\documentclass[journal]{IEEEtran}
\usepackage{cite}
\usepackage{amsmath,amssymb,amsfonts}
\usepackage{graphicx}
\usepackage{textcomp}
\usepackage{xcolor}
\usepackage[utf8]{inputenc}
\usepackage{amsmath}
\usepackage[english]{babel}
\usepackage{amsthm}
\usepackage{multirow}
\usepackage{mathtools}
\usepackage[ruled,norelsize]{algorithm2e}
\usepackage{hyphenat}
\newcommand{\stirlingii}{\genfrac{\{}{\}}{0pt}{}}
\usepackage{url}
\theoremstyle{definition}
\newtheorem{definition}{Definition}
\newtheorem{example}{Example}

\DeclarePairedDelimiter\ceil{\lceil}{\rceil}
\DeclarePairedDelimiter\floor{\lfloor}{\rfloor}

\def\BibTeX{{\rm B\kern-.05em{\sc i\kern-.025em b}\kern-.08em
    T\kern-.1667em\lower.7ex\hbox{E}\kern-.125emX}}
 \newtheorem*{remark}{Remark}

\makeatletter
\newcommand{\removelatexerror}{\let\@latex@error\@gobble}
\makeatother

\begin{document}

\title{Set Partition Modulation}

\author{Ferhat Yarkin~\IEEEmembership{Student~Member,~IEEE} and Justin P.~Coon~\IEEEmembership{Senior~Member,~IEEE}
\thanks{F. Yarkin and J. P. Coon are with the Department of Engineering Science, University of Oxford, Parks Road, Oxford, OX1 3PJ, U.K. E-mail: \{ferhat.yarkin and justin.coon\}@eng.ox.ac.uk}
\thanks{This paper was presented in part at the IEEE International Symposium on Personal, Indoor and Mobile Radio Communications (PIMRC), 8-11 September 2019.}
\thanks{The authors wish to acknowledge the support of the Bristol Innovation \& Research Laboratory of Toshiba Research Europe Ltd.}
}

\maketitle

\begin{abstract}
In this paper, a novel modulation scheme called set partition modulation (SPM) is proposed. In this scheme, set partitioning and ordered subsets in the set partitions are used to form codewords. We define different SPM variants and depict a practical model for using SPM with orthogonal frequency division multiplexing (OFDM). For the OFDM-SPM schemes, different constellations are used to distinguish between different subsets in a set partition. To achieve good distance properties as well as better error performance for the OFDM-SPM codewords, we define a codebook selection problem and formulate such a problem as a clique problem in graph theory. In this regard, we propose a fast and efficient codebook selection algorithm. We analyze error and achievable rate performance of the proposed schemes and provide asymptotic results for the performance. It is shown that the proposed SPM variants are general schemes, which encompass multi-mode OFDM with index modulation (MM-OFDM-IM) and dual-mode OFDM with index modulation (DM-OFDM-IM) as special cases. It is also shown that OFDM-SPM schemes are capable of exhibiting better error performance and improved achievable rate than conventional OFDM, OFDM-IM, DM-OFDM-IM, and MM-OFDM-IM.   
\end{abstract}


\section{Introduction}

The idea of embedding information in the element permutations of a codeword was first proposed by Slepian in \cite{Slepian65}. Slepian's idea, which he called \textit{permutation modulation} (PM), was hinged upon constructing a codebook by permuting elements of a codeword. \textit{Index modulation} (IM) techniques, which can be considered as a subclass of PM, have attracted remarkable interest due to their capabilities for achieving better error performance and improved energy/spectral efficiency compared to conventional systems \cite{Basar2016}. IM encodes the information in the indices of active/inactive sources. For example, combinations of the (in)active transmit antenna indices form the IM codewords in a multi-antenna communication technique called \textit{spatial modulation} (SM) \cite{Mesleh2008}. In contrast,  the activation patterns of the \textit{subcarriers} are used to construct IM codewords in a multicarrier communication technique called orthogonal frequency-division multiplexing with index modulation (OFDM-IM) \cite{Basar2013, Abu2009,Tsonev2011}. Other PM/IM-based applications exist with manifestations in space, time and frequency (see, e.g.,~\cite{Ishikawa2018}). 

The application of IM to the well-known OFDM structure brings various advantages \cite{Basar2013, Abu2009,Tsonev2011, Choi2017,Choi20172, Basar2015, Fan2015, Wen2016, Ishikawa2016, Siddiq2017, Coon2019, Chafii2017}.  For example, it was shown by \cite{Basar2013, Abu2009,Tsonev2011} that OFDM-IM is capable of achieving substantially better error performance than conventional OFDM. To further improve the error performance, transmit diversity  and trellis-coded modulation were applied to OFDM-IM in \cite{Choi2017} and \cite{Choi20172}, respectively. These studies show how one can achieve a diversity gain for the OFDM-IM scheme by sacrificing the data rate. In \cite{Basar2015}, a coordinate interleaved OFDM-IM scheme was proposed, and it was shown that the proposed scheme can achieve an additional diversity gain without sacrificing the data rate. In \cite{Fan2015}, two different generalized OFDM-IM schemes were proposed to improve the spectral efficiency and error performance. Moreover, it was revealed by \cite{Wen2016,Ishikawa2016} that OFDM-IM is capable of achieving a better achievable rate than conventional OFDM for small modulation orders. A useful guideline to design spectrally efficient OFDM-IM schemes is reported in \cite{Siddiq2017}. In \cite{Coon2019}, a binary tree encoding method was applied to OFDM-IM to cover all of the subcarrier activation patterns and, therefore, improve the data rate beyond recent benchmarks. For the same purpose, a discrete cosine transform based solution for OFDM-IM was implemented in \cite{Chafii2017}; substantial data rate improvements compared to OFDM-IM and conventional OFDM were shown possible.              

To encode information in the combinations of the subcarriers, a certain number of subcarriers are nulled in OFDM-IM. Although carrying information on the combinations of active/inactive subcarriers results in better error performance and improved spectral/energy efficiency for small modulation orders, it becomes difficult to achieve spectral efficiencies comparable to conventional OFDM for high modulation orders. To overcome this problem, the idea of employing distinguishable constellations on different subcarriers rather than nulling them has been considered~\cite{Mao2017,Zhang2017,Mao20173,Li2017,Mao20172,Wen2017,Wen2018,Mao2019, Li22017}. In \cite{Mao2017}, an OFDM scheme called \textit{dual mode OFDM with IM} (DM-OFDM-IM) was documented; this method uses two distinguishable constellations rather than active/inactive subcarriers to encode information. In \cite{Zhang2017}, a dual-mode index modulation aided OFDM scheme that employs two PSK constellations with different power levels is proposed. The authors of \cite{Mao20173} further considered altering the number of subcarriers modulated by the same constellation and provided a more general IM scheme called \textit{Generalized DM-OFDM-IM} (GDM-OFDM-IM). In \cite{Li2017}, the authors proposed two different precoding-aided OFDM-IM schemes that use distinguishable constellations. The first scheme of \cite{Li2017} can be considered as a generalization of DM-OFDM-IM since such a scheme partitions the subcarriers into groups and uses the same amount of constellations as the number of groups to modulate the subcarriers.  In \cite{Mao20172}, in addition to two distinguishable constellations, some subcarriers are allowed to remain unused, which yields a third mode that, as it turns out, enhances the bit-error-rate (BER) performance relative to DM-OFDM-IM and OFDM-IM. In \cite{Wen2017}, Wen et al.~proposed a \textit{multi-mode OFDM-IM} (MM-OFDM-IM) scheme, which uses distinguishable constellations on each subcarrier of OFDM sub-blocks to increase the spectral efficiency as well as to improve BER performance. In \cite{Wen2018} and \cite{Mao2019}, the MM-OFDM-IM scheme is generalized. In \cite{Li22017}, the MM-OFDM-IM scheme is extended to space and time domains by using Latin matrices.        

Against this background, we develop a new codebook design method, which we call \textit{set partition modulation} (SPM). Our novel contributions can be listed as follows: 
\begin{itemize}
\item The proposed method uses a novel combinatorial tool, which is partitions of codeword elements rather than permutations or combinations of such elements. 
\item  We define different variants of SPM and give a practical model for applying SPM with OFDM transmissions.
\item The proposed OFDM-SPM schemes employ distinguishable constellations, similar to MM-OFDM-IM; however, in OFDM-SPM, the different constellations are used to distinguish between subsets in a set partition, unlike MM-OFDM-IM where distinguishable constellations are used to construct permutations. 
\item The proposed OFDM-SPM variants are capable of encompassing DM-OFDM-IM and MM-OFDM-IM schemes as special cases. Moreover, they are also capable of achieving a better data rate than such schemes. 
\item We define a codebook selection problem for OFDM-SPM variants to design efficient codebooks, which are at least as good as OFDM-IM schemes in terms of error performance at high SNR. We formulate such a problem as a clique problem and develop an efficient solution for such a problem. 
\item The achievable data rate and BER of OFDM-SPM variants are investigated in this paper, and an upper-bound on the BER is obtained. We also provide asymptotic expressions for data rate and BER. 
\item Our analytical findings show that OFDM-SPM variants are capable of outperforming conventional OFDM, OFDM-IM, DM-OFDM-IM, and MM-OFDM-IM in terms of data rate and BER.
\end{itemize}

The rest of the paper is organized as follows. In Section \ref{section:section2}, we define SPM and its variants. OFDM-SPM and an efficient codebook selection algorithm are described in Sections \ref{section:section3} and \ref{section:section3a}, respectively.  Performance analysis is undertaken in Section \ref{section:section4}. We present and compare analytical and numerical results in Section \ref{section:section5}. Finally, we conclude the paper in Section \ref{section:section6}.

\section{Set Partition Modulation}\label{section:section2}

In this section, we describe the basic idea of SPM.  We begin with some useful definitions and relations. 

\theoremstyle{definition}
\begin{definition}{\textbf{Set Partition:}}
A set partition is the grouping the elements of a set in a way that the groups are disjoint and the union of the groups gives the set.
\end{definition}

\theoremstyle{definition}
\begin{definition}{\textbf{Stirling Number of the Second Kind:}}
The Stirling number of the second kind, denoted by $\stirlingii{N}{K}$, can be defined as the number of ways to partition an $N$-element set into $K$ non-empty subsets. 
\end{definition}

\theoremstyle{definition}
\begin{definition}{\textbf{Bell Number:}}
The Bell number $B_N$ enumerates the total number of partitions of a set of $N$ elements. The Bell number is related to Stirling numbers of the second kind through the equation $B_N=\sum_{K=1}^N \stirlingii{N}{K}$.
\end{definition}

\theoremstyle{definition}
\begin{definition}{\textbf{Ordered Bell Number:}}
The ordered Bell number $\Breve{B}_N$ enumerates the total number of partitions of an $N$-element set considering all permutations of subsets for each partition. The ordered Bell number satisfies the equation $\Breve{B}_N=\sum_{K=1}^N K!\stirlingii{N}{K}$.
\end{definition}

\subsection{SPM}
In an SPM system, a codebook of $L$ codewords $\textbf{x}_1, \textbf{x}_2, \ldots,  \textbf{x}_{L}$ is constructed such that each codeword is a sequence of $N$ elements, which are drawn from a constellation diagram in the complex plane, i.e., $\textbf{x}_l=\big\{x_{l1}, x_{l2}, \ldots, x_{lN}\big\}$, $l=1, \ldots, L$,  where $x_{ln} \in \mathbb C$, $n\in \big\{1,\ldots,N\big\}$, is an $M$-ary symbol. Each codeword is mapped to a partition of an $N$-element set $\mathcal{X}$ into $K$ subsets. Since the number of ways one can form the partition is $\stirlingii{N}{K}$, the SPM codebook size is given by $L_{SPM}=\stirlingii{N}{K}$.   We call $\mathcal{X}$ the generator set of an SPM codeword.  To produce unique codewords for SPM, an $N$-element codeword should have at least $K$ distinguishable elements, and each distinguishable element, $\mu_k, k=1, \dots,K$, is used to specify which element in the codewords belongs to which subset. Therefore, distinguishable elements  differentiate each subset from the other(s). 

\theoremstyle{definition}
\begin{example}
As an example, consider the ways to partition a four-element set $\mathcal{X}\coloneqq\big\{x_{1}, x_{2}, x_{3}, x_{4}\big\}$ into two-element subsets.  This is shown along with the corresponding SPM codewords in Table~\ref{table:table1}. As seen from the table, to obtain the codewords in an SPM codebook, we first partition the elements of the generator set $\mathcal{X}$ into two-element subsets $\mathcal{S}_i$, $i=1,\ldots,L$ where $L=7$\footnote{Note that $\stirlingii{4}{2}=7$.}. Then, we use the subset identifiers $\mu_1$ and $\mu_2$ to represent the elements that belong to different subsets. For example, for the first codeword in Table~\ref{table:table1}, we have the partition $\mathcal{S}_1\coloneqq\big\{\big\{x_{1}\big\}, \big\{x_{2},x_{3}, x_{4}\big\}\big\}$. Since the element $x_1$ and the elements $x_2$, $x_3$, and $x_4$ are in the first and second subsets, respectively, we assign $\mu_1$ to first element and $\mu_2$ to remaining elements. 
\end{example}

\begin{table}[t]
\centering
\caption{Set partitioning and SPM codeword generation example for $N=4$, $K=2$.}
\label{table:table1}
\begin{tabular}{|c|c|}
\hline
\begin{tabular}[c]{@{}c@{}}Set Partitions  of $\mathcal{X}$\end{tabular} & SPM Codeword \\ \hline
$\mathcal{S}_1=\big\{\big\{x_{1}\big\}, \big\{x_{2},x_{3}, x_{4}\big\}\big\}$ & $\textbf{x}_1=\big\{\mu_1, \mu_2, \mu_2, \mu_2\big\}$ \\ \hline
$\mathcal{S}_2=\big\{\big\{x_{2}\big\}, \big\{x_{1},x_{3}, x_{4}\big\}\big\}$ & $\textbf{x}_2=\big\{\mu_2, \mu_1, \mu_2, \mu_2\big\}$ \\ \hline
$\mathcal{S}_3=\big\{\big\{x_{3}\big\}, \big\{x_{1},x_{2}, x_{4}\big\}\big\}$ & $\textbf{x}_3=\big\{\mu_2, \mu_2, \mu_1, \mu_2\big\}$ \\ \hline
$\mathcal{S}_4=\big\{\big\{x_{4}\big\}, \big\{x_{1},x_{2}, x_{3}\big\}\big\}$ & $\textbf{x}_4= \big\{\mu_2, \mu_2, \mu_2, \mu_1\big\}$ \\ \hline
$\mathcal{S}_5=\big\{\big\{x_{1},x_{2}\big\}, \big\{x_{3}, x_{4}\big\}\big\}$ & $\textbf{x}_5=\big\{\mu_1, \mu_1, \mu_2, \mu_2\big\}$ \\ \hline
$\mathcal{S}_6=\big\{\big\{x_{1},x_{3}\big\}, \big\{x_{2}, x_{4}\big\}\big\}$ & $\textbf{x}_6=\big\{\mu_1, \mu_2, \mu_1, \mu_2\big\}$ \\ \hline
$\mathcal{S}_7=\big\{\big\{x_{1},x_{4}\big\}, \big\{x_{2}, x_{3}\big\}\big\}$ & $\textbf{x}_7=\big\{\mu_1, \mu_2, \mu_2, \mu_1\big\}$ \\ \hline
\end{tabular}
\end{table}

\subsection{Ordered SPM} Ordered SPM (OSPM) is an extended version of SPM in which the codebook size is increased by considering permutations of subsets in a partition. Hence, the codebook size of OSPM is given by $L_{OSPM}=K!\stirlingii{N}{K}$.  This simple extension is best illustrated with an example.

\theoremstyle{definition}
\begin{example}
In Table~\ref{table:table2}, we give an example of a mapping between set partitions and OSPM codewords for $N=3$ and $K=2$. As seen from the table, the first three codewords are SPM codewords and we further  obtain an extended codebook by taking the permutations of the subsets in a set partition into account. For this example, by exhibiting additional codewords obtained from the order of the subsets, we end up with $2!\stirlingii{3}{2}=6$ codewords.
\end{example}

\begin{table}[t]
\centering
\caption{Set partitioning and OSPM codeword generation example for $N=3$, $K=2$.}
\label{table:table2}
\begin{tabular}{|c|c|}
\hline
\begin{tabular}[c]{@{}c@{}}Set Partitions  of $\mathcal{X}$\end{tabular} & SPM Codeword \\  \hline
$\mathcal{S}_1=\big\{\big\{x_{1}\big\}, \big\{x_{2},x_{3}\big\}\big\}$ & $\textbf{x}_1=\big\{\mu_1, \mu_2, \mu_2\big\}$ \\ \hline
$\mathcal{S}_2=\big\{\big\{x_{2}\big\}, \big\{x_{1},x_{3}\big\}\big\}$ & $\textbf{x}_2=\big\{\mu_2, \mu_1, \mu_2\big\}$ \\ \hline
$\mathcal{S}_3=\big\{\big\{x_{3}\big\}, \big\{x_{1},x_{2}\big\}\big\}$ & $\textbf{x}_3=\big\{\mu_2, \mu_2, \mu_1\big\}$ \\ \hline
$\mathcal{S}_4=\big\{\big\{x_{2},x_{3}\big\},\big\{x_{1}\big\}\big\}$ & $\textbf{x}_4= \big\{\mu_2, \mu_1, \mu_1\big\}$ \\ \hline
$\mathcal{S}_5=\big\{\big\{x_{1},x_{3}\big\},\big\{x_{2}\big\}\big\}$ & $\textbf{x}_5=\big\{\mu_1, \mu_2, \mu_1\big\}$ \\ \hline
$\mathcal{S}_6=\big\{ \big\{x_{1},x_{2}\big\},\big\{x_{3}\big\}\big\}$ & $\textbf{x}_6=\big\{\mu_1, \mu_1, \mu_2\big\}$ \\ \hline
\end{tabular}
\end{table}

\subsection{Full SPM} In Full SPM (FSPM), the codewords $\textbf{x}_1, \textbf{x}_2, \ldots,  \textbf{x}_{L_{FSPM}}$  are generated by partitioning an $N$-element set $\mathcal{X}$ into non-empty disjoint subsets in such a way that the number of these subsets takes any possible value, $K\in \big\{1, \ldots, N \big\}$. In other words, all partitions of an $N$-element set $\mathcal{X}$ into non-empty disjoint subsets are used to form the FSPM codebook. For FSPM, the codebook size is equal to the Bell number $B_N$, i.e., $L_{FSPM}=B_N=\sum_{K=1}^{N}\stirlingii{N}{K}$. 

\theoremstyle{definition}
\begin{example}
Let us consider an example of how we define the codewords in an FSPM codebook when $N=3$. Partitions of a three-element set $\mathcal{X}\coloneqq\big\{x_{1}, x_{2}, x_{3}\big\}$ are given in Table~\ref{table:table3} along with the corresponding FSPM codewords. As seen from the table, to obtain the codewords in an FSPM codebook, we first partition the elements of the generator set $\mathcal{X}$ into subsets $\mathcal{S}_i$, $i=1,\ldots,L_{FSPM}$ where $L_{FSPM}=5$. Note that Bell number for $N=3$ is $B_3=5$. Then, we use the subset identifier $\mu_k$, $k\in \big\{1, \ldots, K\big\}$, to represent the elements that belong to the $k$th subset.
\end{example}

\begin{table}[t]
\centering
\caption{Set partitioning and FSPM codeword generation example for $N=3$.}
\label{table:table3}
\begin{tabular}{|c|c|}
\hline
\begin{tabular}[c]{@{}c@{}}Set Partitions  of $\mathcal{X}$\end{tabular} & SPM Codeword \\  \hline
$\mathcal{S}_1=\big\{\big\{x_{1}, x_{2}, x_{3}\big\}\big\}$ & $\textbf{x}_1=\big\{\mu_1, \mu_1, \mu_1\big\}$ \\ \hline
$\mathcal{S}_2=\big\{\big\{x_{1}\big\}, \big\{x_{2},x_{3}\big\}\big\}$ & $\textbf{x}_2=\big\{\mu_1, \mu_2, \mu_2\big\}$ \\ \hline
$\mathcal{S}_3=\big\{\big\{x_{2}\big\}, \big\{x_{1},x_{3}\big\}\big\}$ & $\textbf{x}_3=\big\{\mu_2, \mu_1, \mu_2\big\}$ \\ \hline
$\mathcal{S}_4=\big\{\big\{x_{3}\big\}, \big\{x_{1},x_{2}\big\}\big\}$ & $\textbf{x}_4=\big\{\mu_2, \mu_2, \mu_1\big\}$ \\ \hline
$\mathcal{S}_5=\big\{\big\{x_{1}\big\}, \big\{x_{2}\big\}, \big\{x_{3}\big\}\big\}$ & $\textbf{x}_5= \big\{\mu_1, \mu_2, \mu_3\big\}$ \\ \hline
\end{tabular}
\end{table}

\subsection{Ordered Full SPM} We can further increase the number of codewords in an FSPM codebook by considering the permutations of the subsets in a partition. In this regard, we define ordered full SPM (OFSPM) as a modulation scheme that forms its codebook by using all partitions of an $N$-element set $\mathcal{X}$ along with all permutations of the subsets in each partition. Hence, the OFSPM codebook size is given by the ordered Bell number $\Breve{B}_N$, i.e., $L_{OFSPM} = \Breve{B}_N =\sum_{K=1}^{N}K!\stirlingii{N}{K}$.


\section{Practical Model for OFDM}\label{section:section3}

\begin{figure*}[t]
		\centering
	    \resizebox*{12cm}{5.5cm}{\includegraphics{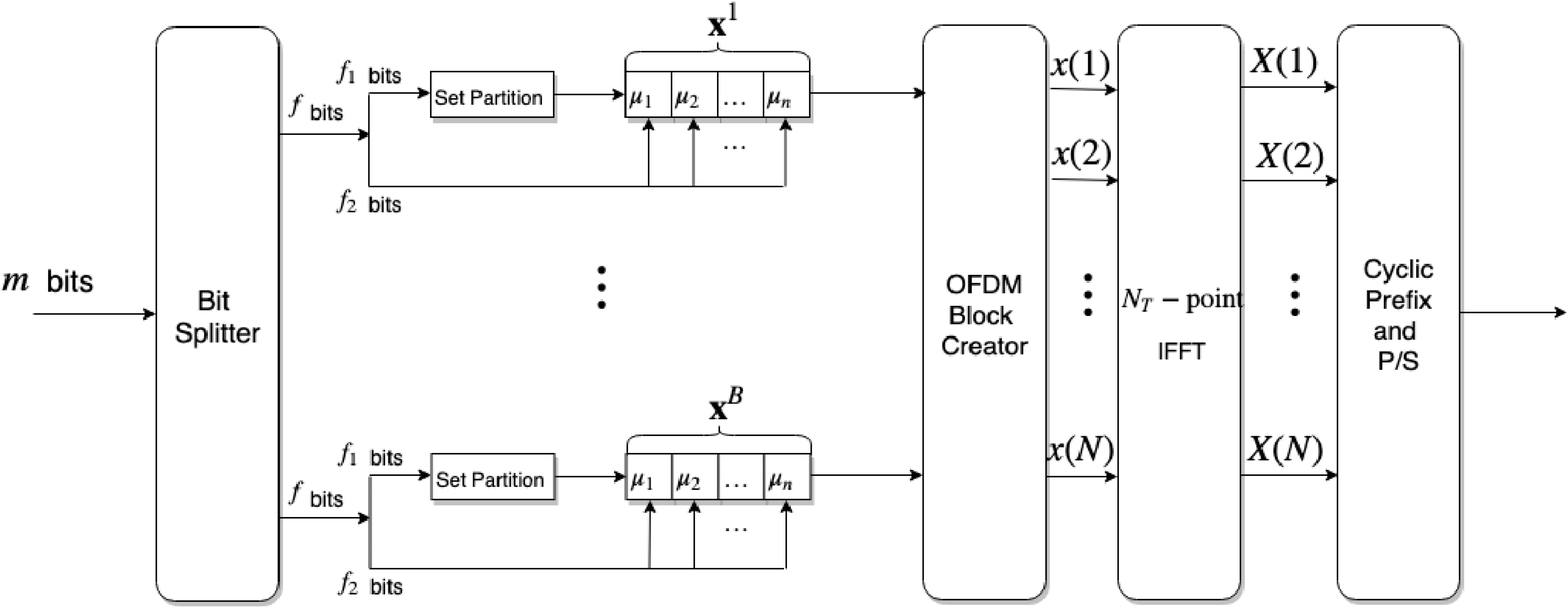}}
		\caption{Transmitter structure of OFDM-SPM scheme.}
		\label{fig:fig1}
\end{figure*}

We present a practical system model in which we apply SPM schemes to OFDM transmissions. The transmitter structure of the OFDM-SPM scheme is shown in Fig.~\ref{fig:fig1}. In this scheme, $m$ input bits enter the SPM transmitter, and these bits are divided into $B=m/f$ blocks, each having $f$ input bits. Similarly, the total number of subcarriers $N_T$ is also divided into $B=N_T/N$ blocks, each having $N$ subcarriers. For each block of input bits, $f$ information bits are modulated by an SPM encoder and the resulting modulated symbols are carried by $N$ subcarriers.    

Since each bit and each subcarrier block have the same mapping operation, we focus on a single block, the $b$th block (where $b\in \big\{1, 2, \ldots, B\big\}$), in what follows. In the $b$th block, the SPM encoder further divides $f$ information bits into two parts, one of them having $f_1$ bits and the other one having $f_2$ bits with $f_1+f_2=f$. The first $f_1$ bits are used to determine the specific set partition $\mathcal{S}_i^b$, $i=1,\ldots,L$ $(L\in \big\{L_{SPM}, L_{OSPM}, L_{FSPM}, L_{OFSPM}\big\})$, of the $N$-element generator set $\mathcal{X}\coloneqq\big\{x_1, x_2, \ldots, x_N\big\}$ belonging to  one of the variants of SPM defined above.  The chosen partition is mapped to the corresponding SPM codeword $\textbf{x}_i^b$ where the superscript $b$ stands for the $b$th block. Here, each element in the SPM codeword corresponds to a subcarrier in the $b$th block. The remaining $f_2$ bits are used to modulate symbols on the $N$ subcarriers, considering the corresponding mapping of the set partition determined by the first $f_1$ bits. As discussed in the previous subsection, we use different subset identifiers in order to produce unique SPM codewords. To preserve the uniqueness property of SPM codewords and modulate the symbols on each subcarrier, we further assume that each subset identifier $\mu_k$ is an element of a disjoint $M$-ary signal constellation $\mathcal{M}_k$, i.e., $\mu_k \in \mathcal{M}_k$ and $\mathcal{M}_k \cap \mathcal{M}_{\hat{k}}=\emptyset$, where $k,\hat{k}\in \big\{1, 2, \ldots, N\big\}$ and $k \neq \hat{k}$\footnote{Note that the subset identifiers are not necessarily elements of disjoint $M$-ary signal constellations and each of them may be chosen as a single constellation point in the same $M$-ary constellation. Hence, assigning a unique constellation point to each subset identifier would be enough to constitute an SPM scheme. However, in this special case, the number of information bits transmitted by an OFDM block is decreased by $f_2$ bits since $f_2$ bits are not used to modulate the subset identifiers.}. For convenience, we choose the size of each constellation as $M$ and, therefore, $f_2=N\log_2 M$. By following the useful design guidelines in \cite{Wen2017}, we obtain the distinguishable PSK constellations $\mathcal{M}_k$ by rotating each constellation with the angle of $2(k-1)\pi/(M N)$, $k=1, \ldots,N$, to maximize the distance between constellation points. To obtain distinguishable QAM constellations, likewise \cite{Wen2017}, we employ the well-known set partitioning technique in \cite{Ungerboeck1982}.

The mapping of $f_1$ bits to the set partitions can be performed by using a look-up table in a similar manner as was proposed to map information bits to subcarrier activation patterns in \cite{Basar2013}, or to permutation indices as detailed in \cite{Wen2017}. A look-up table example illustrating the mapping of $f_1$ bits to the set partitions is given in Table~\ref{table:table4} for $N=4$ and $K=2$. Note that we are only able to use $2^{f_1}$ set partitions.\footnote{It is possible to utilize all set partitions by employing binary coding algorithms \cite{Wang2017,Coon2019}.} As seen from the table, $f_1$ bits are used to determine the specific set partition at first.   The chosen set partition is then used to determine the SPM codeword as discussed earlier.          

\begin{table}[t]
\centering
\caption{A look-up table example corresponding to a bit-to-partition mapping for SPM ($N=4$ and $K=2$). }
\label{table:table4}
\begin{tabular}{|c|c|}
\hline
$f_1$ bits & Set Partitions of $\mathcal{X}$  \\ \hline
{[}0 0{]} & $\mathcal{S}_1^b=\big\{\big\{x_{1}\big\}, \big\{x_{2},x_{3}, x_{4}\big\}\big\}$  \\ \hline
{[}0 1{]} & $\mathcal{S}_2^b=\big\{\big\{x_{2}\big\}, \big\{x_{1},x_{3}, x_{4}\big\}\big\}$  \\ \hline
{[}1 0{]} & $\mathcal{S}_3^b=\big\{\big\{x_{3}\big\}, \big\{x_{1},x_{2}, x_{4}\big\}\big\}$ \\ \hline
{[}1 1{]} & $\mathcal{S}_4^b=\big\{\big\{x_{4}\big\}, \big\{x_{1},x_{2}, x_{3}\big\}\big\}$ \\ \hline
\end{tabular}
\end{table} 

Once the mapping between $f_1$ bits and set partitions has been completed, $f_2=N\log_2 M$ bits are used to determine the modulated symbols, or in other words subset identifiers, on each subcarrier. Hence, one of the SPM codewords $\textbf{x}^b \in \big\{\textbf{x}_1^b,\ldots,\textbf{x}_L^b\big\}$ along with the corresponding modulation symbols $\{\mu_k\}$ constitutes the symbol vector of the $b$th block. After obtaining symbol vectors for all blocks, an OFDM block creator forms the overall symbol vector  $\textbf{x}\coloneqq [x(1), x(2), \ldots,x(N_T)]^T=[\textbf{x}^1,\ldots, \textbf{x}^b, \ldots, \textbf{x}^B]^T\in \mathcal{C}^{N_T\times 1}$. Here, we assume that each element of $\textbf{x}$ is distributed among equally spaced subcarriers to ensure diversity in frequency, and each modulated symbol carried by a subcarrier has unit energy, i.e., $\operatorname{E}[|x(t)|^2]=1$, $t=1,\ldots,N_T$. After this point, exactly the same operations as conventional OFDM are applied.  The symbol vector is processed with an $N_T$-point IFFT, and a cyclic prefix of sufficient length, which is not lower than the memory of the discrete channel impulse response, is attached to the beginning of each time-domain symbol vector. After  parallel-to-serial and up-conversion, transmission is operated over a frequency-selective Rayleigh fading channel.

At the receiver, the received signal is down converted and the cyclic prefix is then removed from each received baseband symbol vector before processing with an FFT. After employing an $N_T$-point FFT operation, the frequency domain received signal vector can be written as 
\begin{align}
    \textbf{y} \coloneqq [y(1),y(2), \ldots, y(N_T)]^T=\sqrt{E_S}\textbf{X}\textbf{h}+\textbf{n}
\end{align}
where $E_S$ is the energy of the transmitted symbol vector and $\textbf{X}=\text{diag}(\textbf{x})$. Moreover, $\textbf{h}$ and $\textbf{n}$ are $N_T\times 1$ channel and noise vectors, respectively. Elements of these vectors follow the complex-valued Gaussian distributions $\mathcal{CN}(0,1)$ and $\mathcal{CN}(0,N_0)$, respectively, where $N_0$ is the noise variance.

Since the encoding procedure for each block is independent of others, decoding can be performed independently at receiver. Hence, using maximum likelihood (ML) detection, the detected symbol vector for the $b$th block can be written as 
\begin{align}\label{eq:eq2}
    \hat{\textbf{x}}^b= \arg \min_{\mathcal{S}_i,\mu_k} ||\textbf{y}^b-\sqrt{E_S}\textbf{X}^b\textbf{h}^b||^2
\end{align}
where $\textbf{y}^b=[y((b-1)N+1), \ldots, y(bN)]^T$, $\textbf{X}^b=\text{diag}(\textbf{x}^b)$ and $\textbf{h}^b=[h((b-1)N+1), \ldots, h(bN)]^T$.

\section{Codebook Selection}\label{section:section3a}
As reported by several studies \cite{Tarokh1998, Wen2017}, for a fading channel, the BER performance of a codebook at a high signal-to-noise ratio (SNR) is limited by the minimum Euclidean distance between codeword pairs whose difference matrix has the minimum rank, i.e., the performance at high SNR is determined by 
\begin{align}
 d_{min}=\min_{i\neq j, i,j \in \big\{1,\ldots,2^f\big\}}||\textbf{X}^i-\textbf{X}^j||^2   
\end{align}
where $\textbf{X}^i=\text{diag}(\textbf{x}^i)$ and $\textbf{X}^j=\text{diag}(\textbf{x}^j)$. Note that here $\text{rank}(\textbf{X}^i-\textbf{X}^j)$ corresponds to the minimum rank among all $\textbf{X}^i$ and $\textbf{X}^j$ pairs. For OFDM-IM schemes, the minimum rank codeword pairs are formed by modulation symbols instead of the symbols carried by the subcarrier indices \cite{Wen2017}. It is fairly easy to see that such minimum rank between codeword pairs is observed when the matrices $\textbf{X}^i$ and $\textbf{X}^j$ regarding OFDM-IM codewords have the same subcarrier indices and different modulation symbols on one of their subcarriers. Moreover, the rank of the difference matrix of two matrices corresponding to different subcarrier indices is at least two \cite{Wen2017}.  Although the statement on the modulation symbols is valid for OFDM-SPM schemes, the statement on symbols carried by subcarrier indices is not necessarily valid for OFDM-SPM schemes. This can be shown by comparing the codewords $\textbf{x}_1$ and $\textbf{x}_5$ in Table \ref{table:table1}. Here, the only difference between the two partitions is observed in the second elements. Such a condition limits the diversity order of the index symbols to one if we include both codewords in the same codebook. However, one can also construct the codebook in a way that unit rank codeword pairs are not included. This ensures the minimum rank properties for OFDM-SPM codewords are the same as for OFDM-IM and MM-OFDM-IM. Hence, our aim in codebook selection for OFDM-SPM variants can be summarized as obtaining codewords that are at least as good as OFDM-IM and MM-OFDM-IM. We know that the optimum codebook selection criterion in a fading channel at high SNR is the well-known rank-determinant criterion \cite{Tarokh1998}. However, due to the complexity of the codebook selection problem, our selection algorithms do not consider product-distances.     

The codebook selection problem can be formulated as a \emph{maximum clique problem}\footnote{This problem is also equivalent to some other important graph problems such as the \emph{maximum independent set problem} and \emph{minimum vertex cover problem.}} or \emph{k-clique problem}  in a graph. Before formulating such problem, we begin with some definitions and notations related to graph theory for clarity.  Consider an undirected graph $G=(V,E)$, where $V=\big\{v_1, v_2, \ldots, v_L\big\}$ is the vertex set and $E \subseteq V \times V$ is the edge set of $G$.  A graph is complete if all its vertices are connected by an edge, i.e., we have $(v_l, v_{\hat{l}}) \in E$ for $\forall v_l, v_{\hat{l}} \in V$, $v_l\neq v_{\hat{l}}$. A clique can be defined as a complete subgraph of a graph \cite{CARRAGHAN1990, Bomze1999}. The clique number or the size of the maximum clique is denoted by $w(G)$. Moreover, the symmetric $L \times L$ matrix $A_G=(a_{v_l v_{\hat{l}}})_{{v_l, v_{\hat{l}}} \in V}$, where $a_{v_l v_{\hat{l}}}=1$ if $(v_l, v_{\hat{l}})\in E$ and $a_{v_l v_{\hat{l}}}=0$ if $(v_l, v_{\hat{l}})\notin E$, stands for the adjacency matrix of the graph $G$.  

Now, we formulate our codebook selection problem as maximum clique and $k$-clique problems as follows. Assume that a graph's vertices represent the OFDM-SPM codewords, which map to set partitions, and the graph's edges represent the Hamming distance between such codeword pairs\footnote{Note that the Hamming distance between $\textbf{x}_i$ and $\textbf{x}_j$ is equivalent to the rank of the difference matrix $\textbf{X}^i-\textbf{X}^j$ where $\textbf{X}^i=\text{diag}(\textbf{x}_i)$ and $\textbf{X}^j=\text{diag}(\textbf{x}_j)$. Here, we prefer to use Hamming distance rather than rank for convenience.}. To ensure similar Hamming distance properties for index symbols as OFDM-IM benchmarks, we set a condition on the vertex connections such that two vertices are connected to each other if the Hamming distance between vertices (codewords) is greater or equal to two, i.e. $E=\big\{(v_l, v_{\hat{l}})|~ v_l, v_{\hat{l}} \in V,  \operatorname{HamDist} (v_l, v_{\hat{l}})\ge 2\big\}$ where the ``HamDist'' function is used to measure the Hamming distance between vertices. Here, our aim is to finding the maximum clique that includes the largest possible number of vertices. 

On the other hand, if the problem is formulated as a \emph{k-clique problem}, one should consider the cliques that includes a specific number, $k$, of vertices. Considering the number of the set partitions, the clique size, $k$, can be determined. For example, we have $\Breve{B}_N=75$ ordered set partitions for $N=4$ and we are interested in choosing 64 partitions out of 75 to send six data bits. Hence, we can formulate such a problem in the same way as the maximum clique problem. However, in this case, we are interested in the cliques with 64 vertices. One drawback of the $k$-clique problem is that we do not know whether the current graph has a clique of size $k$. However, we can proceed to check the cliques by decreasing the clique size in the absence of a $k$-clique in the given graph.

\subsection{Brute-force Search for $k$-clique Problem}

 One straightforward way of selecting good codewords or, in other words finding a $k$-clique, is exhaustively searching among all possible subgraphs. In this regard, the brute-force search algorithm for the $k$-clique problem is given in Algorithm \ref{alg:alg1}. The algorithm starts by initializing the graph $G(V,E)$ where $V=\big\{v_1, v_2, \ldots, v_L\big\}$ is the vertex set  representing SPM codewords and, $E=\big\{(v_i, v_j)|~ v_i, v_j \in V,  \operatorname{HamDist}(v_i, v_j)\ge 2 \big\}$ is the edge set of $G$. In the second step, we set $\kappa=0$ where $\kappa$ is used to adjust the clique size. Then, in the third step,  we divide the graph constituted by the partitions into subgraphs, $V_i$, in a way that each subgraph includes $k=2^{\floor*{\log_2{w_U(G)}}-\kappa}$ vertices where $w_U(G)=N_{-1}+1$ is an upper-bound on the size of the  maximum clique $w(G)$, i.e., $w(G)\leq w_U(G)=N_{-1}+1$\footnote{One may find many bounds on the clique number \cite{Bomze1999}. The safest strategy would be calculating all bounds and using the tightest. We found the upper-bound in \cite{Amin1972} sufficiently tight for our problem.} \cite{Amin1972}. Here, we choose the number of vertices, $k$, in a subgraph as a power of two since the complexity of the algorithm would be high if we take all possible values of $k$ into consideration. The choice of $k$ can also be justified by the use of a fixed-length bit mapping scheme, since we wish to send the same amount of data bits even if we consider all possible values of $k$. Moreover,   $N_{-1}$ denotes the number of eigenvalues of the adjacency matrix $A_G$ that do not exceed $-1$. Then, for each subgraph, we calculate adjacency matrices, $A_{G_i}$, and initialize a selected graph as $\hat{V}\in \emptyset$. In the fifth step, we check each adjacency matrix to see whether the subgraph is a $k$-clique or not. If the subgraph is a $k$-clique, the algorithm returns the current subgraph as the selected graph. If no $k$-clique is encountered after checking all subgraphs, we increase $\kappa$ by one and go to Step 3 to check whether there is a clique of smaller size. In that way, the brute-force search algorithm guarantees reaching the maximum number of bits carried by set partitions for fixed-length binary coding schemes. Note that subgraph initialization in Step 3 may require excessive memory when the number of vertices is high. To avoid such memory requirements, one can use a combinatorial number system to obtain vertex combinations from natural numbers. The combinatorial number system provides a bijective mapping between the natural numbers and $k$-combinations \cite{Knuth1997}. Hence, Steps 3 and 4 can be integrated into the for-loop in Step 5 and each subgraph can be constructed in each loop by using a combinatorial number system. Note also that the for-loop in Step 5 can be parallelized to speed up the algorithm.    

\begin{figure}[t]
 \removelatexerror
  \begin{algorithm}[H]
   \caption{Brute-force search for the $k$-clique problem}
   \label{alg:alg1}
   {\bf Step 1:} Initialize a graph $G(V,E)$ with $V=\big\{v_1, v_2, \ldots, v_L\big\}$, the vertex set  of SPM codewords, and, $E=\big\{(v_l, v_{\hat{l}})|~ v_l, v_{\hat{l}} \in V,  \operatorname{HamDist}(v_l, v_{\hat{l}})\ge 2 \big\}$\;
   {\bf Step 2:} $\kappa=0$\;
    {\bf Step 3:} Initialize subgraphs, $G_i(V_i, E)$, $i=1,\ldots, C$, with $V_i=\big\{v_1^i, v_2^i,\ldots, v_{k}^i \big\}$ where $C=\binom{L}{k}$, $k=2^{\floor*{\log_2 w_U(G)}-\kappa}$, $v_{l}^i, v_{\hat{l}}^i \in V$, $l,\hat{l}=\big\{1, 2, \ldots, k\big\}$, $v_{l}^i\neq v_{\hat{l}}^i$\;
    {\bf Step 4:} Initialize adjacency matrix, $A_{G_i}$, of $G_i$ and $\hat{V}\in\emptyset$ where $A_{G_i}=(a_{l, \hat{l}}^i)_{(v_{l}^i, v_{\hat{l}}^i)\in V_i}$\;
    {\bf Step 5:} Finding the $k$-clique\; 
    \For{$i=1$ \KwTo $C$}
    {
        {\If{$a_{l, \hat{l}}^i=1$, $\forall l, \hat{l} = \big\{1, 2, \ldots, k\big\}$ \text{and} $l\neq \hat{l}$}
        {$\hat{V}=V_i$;
        
        {break;}}
        
          }
        $i \leftarrow i+1$\;}
     {\bf Step 6:} Checking the presence of $k$-clique in $G$\; 
     \If{$\hat{V}\in\emptyset$}{
     $\kappa \leftarrow \kappa+1$\; 
     Go to Step 3\;
     }
        {\bf return} $\hat{V}$\;
  \end{algorithm}
\end{figure}

\subsection{Vertex Exclusion for the Maximum Clique Problem}

Since the maximum clique problem is NP-complete, the time of execution of exact algorithms will increase exponentially with the number of vertices in the graph \cite{Bomze1999}. In this regard, the brute-force search algorithm exhibits  high complexity when the number of vertices is high. To overcome this problem, we propose a more practical algorithm based on excluding the vertices that have the minimum number of connections. Although such an algorithm provides a sub-optimal solution for the codeword selection problem, it is much more efficient than a brute-force search in terms of complexity. Our proposed algorithm is given in Algorithm \ref{alg:alg2}. The algorithm starts by initializing the Hamming graph. Then, in the second step, we calculate the number of neighbors, $|N(v_l)|$, of each node $v_l\in V$ where $N(v_l)=\big\{v_{\hat{l}}\in V;~ a_{v_lv_{\hat{l}}}=1\big\}$ is the neighborhood of $v_l$ in $G$. In the third step, we remove the nodes that have the minimum number of neighbors until the remaining graph is a clique, i.e. $|N(v_l)|=|V|-1, \forall{v_l}\in V$. In this step, $|N_c(v_l)|$ is used to update the cardinality of the neighborhood of $v_l$ after removing the node that has the minimum number of neighbors.  As a result, the algorithm removes the set partitions having the maximum number of unit Hamming distance pairs one-by-one until we have no set partition pairs with unit Hamming distances.

\begin{figure}[t]
 \removelatexerror
  \begin{algorithm}[H]
   \caption{Vertex exclusion for maximum clique problem}
   \label{alg:alg2}
   {\bf Step 1:} Initialize a graph $G(V,E)$ with $V=\big\{v_1, v_2, \ldots, v_L\big\}$, the vertex set  of SPM codewords, and, $E=\big\{(v_l, v_{\hat{l}})|~ v_l, v_{\hat{l}} \in V,  \operatorname{HamDist}(v_l, v_{\hat{l}})\ge 2 \big\}$\;
  {\bf Step 2:} Calculate the cardinality of the neighborhood, $|N(v_l)|$, for each node where $v_l\in V$  \;
  {\bf Step 3:} Finding a clique by excluding vertices\;
    \While{$|N(v_l)|\neq|V|-1, \forall v_l\in V$}
    { 
    
    $v_{min}=\arg \min_{v_l\in V}|N(v_l)|$\;
    $G\leftarrow G - \big\{v_{min}\big\}$\;
    $V\leftarrow V - \big\{v_{min}\big\}$\;
    Calculate  the neighborhood cardinality, $|N_c(v_l)|$, for each node of $V$ where $\forall v_l\in V$\;
    $|N(v_l)|\leftarrow |N_c(v_l)|$\;

}

        {\bf return} $V$\;
  \end{algorithm}
\end{figure}

\subsection{Complexity Comparison}
  In this subsection, we compare the complexities of Algorithms \ref{alg:alg1} and \ref{alg:alg2} in terms of the algorithm run-times. Results in units of seconds are shown in Tables \ref{table:table5} and \ref{table:table6}. In the tables, we also depict the numbers of vertices achieved by each algorithm in the brackets along with the upper-bound results on the achievable number of vertices according to \cite{Amin1972}. Moreover,  results using a built-in function called ``FindClique[$G$]'' in Wolfram Language are provided as a benchmark. Such a function searches for the maximal set of vertices where the corresponding subgraph is a clique \cite{Weisstein2019}. 

In Table \ref{table:table5}, the complexity results are provided  for OFDM-OSPM schemes having $N=4, 6$ and, $8$ subcarriers and $K=2$ distinguishable constellations in each OFDM sub-block. In this case, OFDM-OSPM schemes can produce 14, 62 and 254 vertices (SPM codewords) for $N=4, 6$ and, $8$, respectively. On the other hand, in Table \ref{table:table6}, we provide complexity results for OFDM-OFSPM schemes having $N=3, 4, 5$ and $6$ subcarriers. In that case, OFDM-OFSPM schemes are capable of producing 13, 75, 541, and, 4683 vertices (SPM codewords) for $N=3, 4, 5$ and $6$, respectively.  As can be seen from Tables \ref{table:table5} and \ref{table:table6}, Algorithm \ref{alg:alg1} is not able to provide a practical solution for most of the cases\footnote{Here,``-'' denotes that the algorithm is not able to return a solution within a reasonable amount of time.}. However, Algorithm \ref{alg:alg2} provides  efficient solutions for all of the cases by outperforming both Algorithm \ref{alg:alg1} and the ``FindClique[$G$]'' function in terms of the algorithm run-time. The effectiveness of Algorithm \ref{alg:alg2} becomes more evident when the number of  subcarriers, and therefore the number of vertices, increases. Furthermore, the clique number achieved by Algorithm \ref{alg:alg2} is  consistent with the clique numbers achieved by the ``FindClique[$G$]'' function and obtained by the upper-bound in \cite{Amin1972}.

\begin{table*}[t]
\centering
\caption{The run-time complexities (in seconds) of the vertex selection algorithms along with the numbers of achieved vertices (in brackets) for OFDM-OSPM scheme with $N\in \big\{4, 6, 8\big\}$ and $K=2$.}
\label{table:table5}
\begin{tabular}{|c|c|c|c|c|}
\hline
 & Algorithm 1 & Algorithm 2 & FindClique{[$G$]} & Upper-bound in \cite{Amin1972} \\ \hline
$N=4$ & 0.4343 (8) & 0.0033 (8)  & 0.05 (8) & (9) \\ \hline
$N=6$ & - & 0.0176 (32) & 0.06 (32) & (32) \\ \hline
$N=8$ & - & 0.0761 (128)  & 0.20 (128) & (129)  \\ \hline
\end{tabular}
\end{table*}

\begin{table*}[t]
\centering
\caption{The run-time complexities (in seconds) of the vertex selection algorithms along with the numbers of achieved vertices (in brackets) for OFDM-OFSPM scheme with $N\in \big\{3, 4, 5, 6\big\}$.}
\label{table:table6}
\begin{tabular}{|c|c|c|c|c|}
\hline
 & Algorithm 1 & Algorithm 2 & FindClique{[$G$]} & Upper-bound in \cite{Amin1972} \\ \hline
$N=3$ & 0.024 (4) & 0.0022 (7)  & 0.03 (7) & (7) \\ \hline
$N=4$ & - & 0.0088 (32) & 0.06 (32) & (33) \\ \hline
$N=5$ & - & 0.4929 (181)  & - & (225)  \\ \hline
$N=6$ & - & 477.0625 (1321)  & - & (1876)  \\ \hline
\end{tabular}
\end{table*}


\section{Performance Analysis}\label{section:section4}

In this section, we analyze the data rate and the BER of the proposed SPM schemes. 

\subsection{Data Rate}\label{AA}

We analyze the data rate of the proposed OFDM-SPM schemes in terms of the number of bits corresponding to the OFDM-SPM codewords normalized by the number of subcarriers used to convey each codeword. Here, we do not take cyclic prefix length into account for convenience. We also provide some useful expressions for the number of partitions obtained by SPM schemes along with comparisons regarding OFDM-IM benchmarks. We assume $f_2=N\log_2M$ for each SPM variant. 

\subsubsection{OFDM-SPM} As discussed in Section \ref{section:section2}, the number of set partitions produced by SPM is given by the Stirling number of the second kind $\stirlingii{N}{K}$. Considering the fact that each symbol on each subcarrier has the modulation order $M$, the achievable data rate per subcarrier for an OFDM-SPM scheme with $N$ subcarriers and $K$ subset identifiers in each sub-block is
\begin{align}
    R_{SPM}=\frac{f_1+f_2}{N}=\frac{\floor{\log_2\stirlingii{N}{K}}+N\log_2M}{N}
\end{align}
where $\floor{.}$ is the floor operation.

The Stirling number of the second kind, $\stirlingii{N}{K}$, can be written as \cite{Temme93} 
 \begin{align}
 \stirlingii{N}{K}= \frac{1}{K!} \sum_{j=0}^{K}(-1)^j\binom{K}{j}(K-j)^N.  
 \end{align}
It is also straightforward to show that the following recurrence holds:
 \begin{align}\label{eq:eq6}
      \stirlingii{N}{K}=K \stirlingii{N-1}{K}+ \stirlingii{N-1}{K-1}.
 \end{align}
 
\begin{remark}\label{remark:remark1}
Consider an OFDM-SPM block having two distinguishable constellations, i.e., $K=2$, on $N$ subcarriers and a DM-OFDM-IM block having two distinguishable constellations on $N$ subcarriers in which $N-d$ of the subcarriers is modulated by one of two different constellations and the remaining $d$ of them is modulated by the other constellation.\footnote{It is fair to compare these two schemes since both of the schemes has two distinguishable constellations.} Except for the case where $N=2$ and $d=1$, the number of set partitions obtained by the OFDM-SPM encoder is equal to or greater than the number of subcarrier combinations obtained by the DM-OFDM-IM encoder, i.e., $\stirlingii{N}{2}\ge\binom{N}{d}$ for $N\ge2$. The equality holds for $N=2$ and $d=2$. This means that the achievable data rate for OFDM-SPM is equal to or greater than that of DM-OFDM-IM when the subcarriers of both schemes carry symbols that have the same modulation order. One may check that $\stirlingii{N}{2}=2^{N-1}-1$. Moreover,   
from the recurrence relation of the binomial coefficient, we have $\binom{N}{d}=\binom{N-1}{d}+\binom{N-1}{d-1}$. If we compare $\binom{N}{d}$ with $\stirlingii{N}{2}$, we see that   $\stirlingii{N}{2}\ge\binom{N}{d}$ for $N\ge2$ except for the case where $N=2$ and $d=1$.  
\end{remark}

For a fixed and relatively small value of $K$, the asymptotic value of the Stirling number of the second kind as $N\to \infty$ can be written as $\stirlingii{N}{K}\sim \frac{K^N}{K!}$. Hence, the asymptotic value of the achievable data rate per subcarrier for an OFDM-SPM scheme as $N \to \infty$ can be written as 
\begin{align}\label{eq:eq7}
    R_{SPM}&\sim \frac{{N\log_2K-\log_2K!}+N\log_2M}{N}
    \\ \nonumber& \sim \log_2(KM).
\end{align}

\begin{remark}
Consider a special MM-OFDM-IM scheme as in \cite{Li2017} having $N$ subcarriers and $K$ distinguishable constellations in each OFDM sub-block along with $M$-PSK symbols on each subcarrier. To compare the achievable rate of such an MM-OFDM-IM scheme, we assume that each $N/K$ subcarriers employ the same constellation/mode. Hence, for this scenario, we have $\frac{N!}{(N/K)!^K}$ mode combinations. Applying the Stirling approximation\footnote{$\ln N!\sim N\ln N-N$ as $N\to \infty$.}, the achievable data rate per subcarrier for this MM-OFDM-IM scheme as $N\to \infty$ can be written as $R_{MM-OFDM-IM}\sim \log_2(KM)$. Hence, the proposed OFDM-SPM scheme is capable of providing asymptotically the same achievable rate as this special MM-OFDM-IM scheme.       
\end{remark}

 When $N$ is large, the value of $K$ that maximizes $\stirlingii{N}{K}$ satisfies $K_N\sim \frac{N}{\ln N}$ \cite{Rennie69}. More precisely, the following relation holds for sufficiently large $N$ \cite{Canfield2002} \begin{align}\label{eq:eq8-1}
    K_N \in \big\{\floor{e^{W(N)}-1}, \ceil{e^{W(N)}-1} \big\}
\end{align}
where $\floor{.}$ and $\ceil{.}$ are floor and ceiling operations, respectively. $W(N)$ is the Lambert W function satisfying $W(N)\exp(W(N))=N$ \cite{Corless1996}. There is no exception to the relation in \eqref{eq:eq8-1} for $1 \le N \le 1200$  \cite{Canfield2002}. More importantly, the maximum value of the Stirling number of the second kind adheres to the relation
$\ln \stirlingii{N}{K_N}\sim N\ln N-N\ln \ln N-N$ \cite{Rennie69}. Hence, the asymptotic maximum achievable rate of the OFDM-SPM scheme satisfies
\begin{align}\label{eq:eq8}
 R_{SPM}^{max}&\sim \frac{{N\log_2 ({N}/{e\ln N)}}+N\log_2 M}{N} \\ \nonumber& \sim \log_2(N/\ln N)+\log_2 (M)-\log_2 e . 
\end{align}
This asymptotic result shows that a substantially improved data rate is attainable when we use $K_N$ distinguishable constellations in the OFDM-SPM scheme.

 \subsubsection{OFDM-OSPM} The achievable data rate per subcarrier for an OFDM-OSPM scheme having $N$ subcarriers and $K$ subset identifiers in each sub-block can be written as 
 \begin{align}
    R_{OSPM}=\frac{f_1+f_2}{N}=\frac{\floor{\log_2K!\stirlingii{N}{K}}+N\log_2M}{N}
\end{align}
\begin{remark}
It is straightforward to show that $2!\stirlingii{N}{2}\ge\binom{N}{d}$. However, it is important to note that an OFDM-OSPM codebook, which incorporates $K=2$ element partitions and their ordered counterparts, subsumes a DM-OFDM-IM codebook. This can easily be proved by considering set partitions along with permutations when $K=2$. Hence, it can be concluded that OFDM-OSPM is a more general scheme, which encompasses the subcarrier combinations generated by a DM-OFDM-IM encoder. It is also important to note that an OFDM-OSPM encoder produces the same index patterns as an MM-OFDM-IM encoder for $K=N$. Hence, MM-OFDM-IM is a special case of OFDM-OSPM when $K=N$. Moreover, the  partitions of an $N$-element set into two subsets would result in the same index symbols as GDM-OFDM-IM when the set, $\mathcal{K}$, containing possible numbers of subcarriers having one of a number of distinguishable constellations in each OFDM sub-block is given by $\mathcal{K}=\big\{1,2, \ldots, N-1\big\}$ for GDM-OFDM-IM.  \textit{Despite these similarities to IM schemes, it is important to recognize that OSPM is inherently different due to the use of set partitions to encode information instead of index patterns or permutations.}
\end{remark}

Assuming a fixed $K$ and using the asymptotic representation for  the Stirling number of the second kind, we can write the following achievable data rate expression as $N \to \infty$   for OFDM-OSPM:
\begin{align}
    R_{OSPM}&\sim \frac{{N\log_2K}+N\log_2M}{N}
    \\ \nonumber& \sim \log_2(KM).
\end{align}
Hence, OFDM-OSPM achieves the same asymptotic data rate as OFDM-SPM when $K<<N$. On the other hand, it is known that, the value of $K$ that maximizes $K!\stirlingii{N}{K}$ satisfies $\breve{K}_N \sim \frac{N}{2 \ln 2}$ as $N\to \infty$ \cite{mezo2015}. Moreover, we have the asymptotic relation $\Breve{K}_N!\stirlingii{N}{\breve{K}_N} \sim N!/2(\ln 2)^{N+1}$ \cite{mezo2015}. Hence,  the  asymptotic  maximum  achievable  rate  of  the OFDM-OSPM scheme satisfies
\begin{align}\label{eq:eq12-1}
    R_{OFSPM}^{max}&\sim \frac{{\log_2(N!/2(\ln 2)^{N+1})}+N\log_2 M}{N} \\ \nonumber& \sim \log_2(N)+\log_2 (M)-\log_2 (e\ln 2).
\end{align}
where the second asymptotic relation follows from the Stirling's approximation.

We can conclude that the OFDM-OSPM scheme is capable of achieving  asymptotically better achievable rate than the MM-OFDM-IM scheme while the number of exploited distinguishable constellations is less than $N$.

\subsubsection{OFDM-FSPM} The achievable data rate per subcarrier for an OFDM-FSPM scheme having $N$ subcarriers can be written as 
 \begin{align}
    R_{FSPM}=\frac{f_1+f_2}{N}=\frac{\floor{\log_2B_N}+N\log_2M}{N}.
\end{align}

 Using the asymptotic expression for the Bell number given in \cite{Brujin}, the asymptotic achievable data rate of OFDM-FSPM  can be written as
\begin{align}\label{eq:eq14}
    R_{FSPM} \sim \log_2(N/\ln N)+\log_2 (M)-\log_2 e. 
\end{align}

\subsubsection{OFDM-OFSPM} The achievable data rate per subcarrier for an OFDM-OFSPM scheme having $N$ subcarriers can be written as 
 \begin{align}\label{eq:eq9}
    R_{OFSPM}=\frac{f_1+f_2}{N}=\frac{\floor{\log_2\Breve{B}_N}+N\log_2M}{N}.
\end{align}

\begin{remark}
It is  straightforward to show that $\Breve{B}_N>N!$ for $N\ge2$, since the ordered set partitions include all  permutations of $N$-element set partitions and the number of the permutations of $N$-element  partitions is equal to $N!$. In other words, for an $N$-element set $\mathcal{X}$ with $K=N$, $\mathcal{S}=\big\{\big\{x_1\big\}, \big\{x_2\big\}, \ldots, \big\{x_N\big\} \big\}$ is a valid set partition, and ordering the subsets of this set would result in $N!$ different partitions. Hence, the resulting OFDM-OFSPM codebook contains the codeword $\textbf{x}^b=\big\{\mu_1, \mu_2, \ldots, \mu_N\big\}$ along with the codewords representing all  permutations of the elements of $\textbf{x}^b$. Note that these codewords form the MM-OFDM-IM codebook, and OFDM-OFSPM is a more general scheme compared to MM-OFDM-IM since it covers  all codewords formed by an MM-OFDM-IM encoder. 
\end{remark}

Using the asymptotic relation for ordered Bell numbers given in \cite{BARTHELEMY1980}, the asymptotic achievable data rate of OFDM-OFSPM scheme as $N\to \infty$ can be written as
\begin{align}\label{eq:eq17}
    R_{OFSPM}&\sim \frac{{\log_2(N!/2(\ln 2)^{N+1})}+N\log_2 M}{N} \\ \nonumber& \sim \log_2(N)+\log_2 (M)-\log_2 (e\ln 2).
\end{align}

As can be observed from \eqref{eq:eq8}, \eqref{eq:eq12-1}, \eqref{eq:eq14}, and \eqref{eq:eq17}, the maximum achievable data rate of each SPM variant exhibits the same asymptotic behavior as its full SPM counterpart while exploiting fewer distinguishable constellations. Hence, OFDM-SPM and OFDM-OSPM are capable of achieving the same asymtotic date rate as OFDM-FSPM and OFDM-OFSPM, respectively, by utilizing fewer constellation points.

\subsection{Bit-Error Rate}
Let $P(\textbf{X}^i\to\textbf{X}^j)$ denote the pairwise error probability (PEP) associated with the erroneous detection of $\textbf{X}^i$ as $\textbf{X}^j$ where $\textbf{X}^i=\text{diag}(\textbf{x}^i)$ and $\textbf{X}^j=\text{diag}(\textbf{x}^j)$. From \eqref{eq:eq2}, the PEP conditioned on the channel coefficients is given by
\begin{align}
    P(\textbf{X}^i\to\textbf{X}^j|\textbf{h})=Q\Bigg(\sqrt{\frac{E_S||(\textbf{X}^i-\textbf{X}^j)\textbf{h}||^2}{2N_0}}\Bigg).
\end{align}

By using the identity $Q(x)\approx \frac{1}{12}e^{-x^2/2}+\frac{1}{4}e^{-2x^2/3}$ and averaging over $\textbf{h}$, an approximate unconditional PEP expression can be obtained \cite{Basar2013}:
\begin{align}\label{eq:eq12}
    P(\textbf{X}^i\to\textbf{X}^j)&=\operatorname{E}_{\textbf{h}}\big[P(\textbf{X}^i\to\textbf{X}^j|\textbf{h})\big]\nonumber\\ &\approx \frac{1/12}{\det(\textbf{I}_N+\frac{E_S}{4N_0}\textbf{C}\textbf{Z}_{ij})} \nonumber \\
    &\qquad+\frac{1/4}{\det(\textbf{I}_N+\frac{E_S}{3N_0}\textbf{C}\textbf{Z}_{ij})}
\end{align}
where $\textbf{I}_N$ denotes the $N\times N$ identity matrix, $\textbf{C}=\operatorname{E}_{\textbf{h}}[\textbf{h}\textbf{h}^H]$ and $\textbf{Z}_{ij}=(\textbf{X}^i-\textbf{X}^j)^H(\textbf{X}^i-\textbf{X}^j)$.

An upper-bound on the average BER is given by the well-known union bound
\begin{align}\label{eq:eq13}
    P_b \leq \frac{1}{f2^f}\sum_{i=1}^{2^f}\sum_{j=1}^{2^f}P(\textbf{X}^i\to\textbf{X}^j)D(\textbf{X}^i\to\textbf{X}^j)
\end{align}
where $D(\textbf{X}^i\to\textbf{X}^j)$ is the number of bits in error for the corresponding pairwise error event. Note that the upper-bound expression given in \eqref{eq:eq13} is valid for all OFDM-SPM schemes.

Assuming $\textbf{C}=\operatorname{E}_{\textbf{h}}[\textbf{h}\textbf{h}^H]\approx \textbf{I}_N$ and considering the fact that $\textbf{Z}_{ij}$ is a diagonal matrix, one can rewrite \eqref{eq:eq12} as 
\begin{align}\label{eq:eq21}
        P(\textbf{X}^i\to\textbf{X}^j)&\approx \frac{1/12}{\prod_{n=1}^{N}(1+\frac{E_S}{4N_0}\lambda_n)} \nonumber \\
    &\qquad+\frac{1/4}{\prod_{n=1}^{N}(1+\frac{E_S}{3N_0}\lambda_n)}
\end{align}
where $\lambda_n$ is the $n$th diagonal element of $\textbf{Z}_{ij}$. At high SNR, one can neglect the one in the denominator of \eqref{eq:eq21} and write the following approximation 
\begin{align}\label{eq:eq22}
        P(\textbf{X}^i\to\textbf{X}^j)\approx \frac{1/12}{\prod_{n\in \Gamma}\frac{E_S}{4N_0}\lambda_n}+\frac{1/4}{\prod_{n \in \Gamma}\frac{E_S}{3N_0}\lambda_n}
\end{align}
where $\Gamma$ is the set of nonzero diagonal elements of $\textbf{Z}_{ij}$. Finally, an asymptotic expression for the BER of the OFDM-SPM variants can be obtained by substituting  \eqref{eq:eq22} into  \eqref{eq:eq13}. 

As explained in the previous sections, OFDM-SPM codebooks can be designed in a way that the minimum Hamming distance between index symbols for the set partitions is equal to two. Hence, assuming the OFDM-SPM index symbols achieve this minimum Hamming distance property, the average BER expression will be dominated by the modulation symbols at high SNRs and, therefore, the diversity order of the BER curves is limited to one.


\section{Numerical Results}\label{section:section5}
In this section, we provide numerical data rate and BER results for the proposed schemes. In figures, OFDM-SPM $(N,K, M)$ and OFDM-OSPM $(N,K, M)$ stand for OFDM-SPM schemes having $N$ subcarriers and $K$ distinguishable constellations in each OFDM sub-block along with $M$-PSK symbols on each subcarrier, whereas OFDM-FSPM $(N, M)$ and OFDM-OFSPM $(N, M)$ stand for variants of OFDM-SPM employing all set partitions and having $N$ subcarriers with $M$-PSK symbols in each OFDM sub-block. Moreover, OFDM-IM $(N,K_a, M)$ stands for the conventional OFDM-IM scheme having $K_a$ activated subcarriers out of $N$ in each sub-block and employing $M$-PSK modulation on the activated subcarriers. Finally, DM-OFDM-IM $(N, M)$ signifies a dual-mode scheme having $N$ subcarriers along with two distinguishable $M$-PSK constellations, and MM-OFDM-IM $(N, M)$ represents a multi-mode scheme having $N$ subcarriers along with $N$ distinguishable $M$-PSK constellations in each sub-block.

In our simulations, we assume all schemes operate over a Rayleigh fading channel, whose elements are independent and identically distributed.  ML detection is applied under the assumption that channel estimation is perfect. The use of the Rayleigh model in the simulations is realistic when considering environments with a large number of scatterers~\cite{Vucetic2003}.

\subsection{Data Rate}

\begin{figure}[t]
		\centering
        \includegraphics[width=8.5cm]{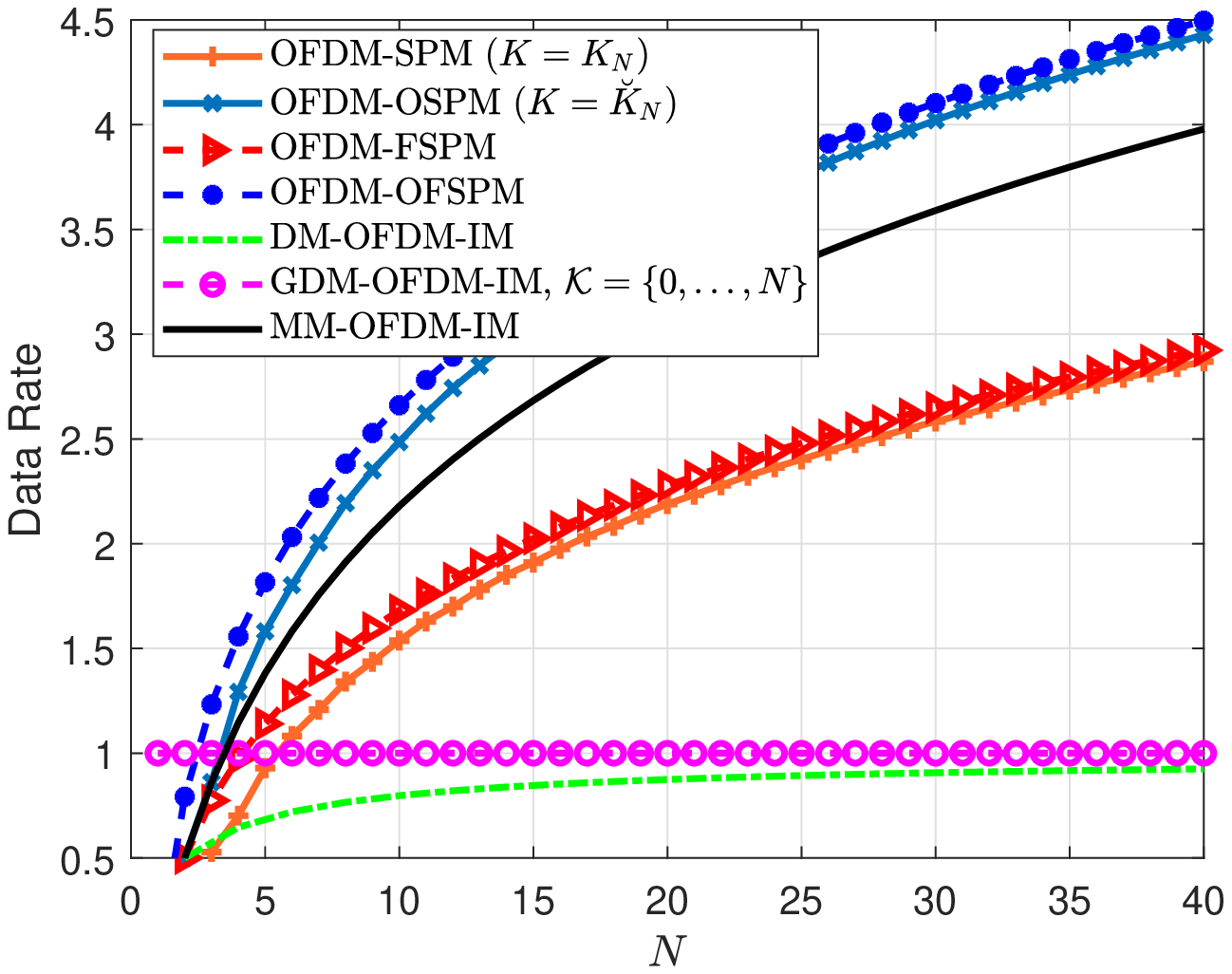}
		\caption{Data rate comparison of OFDM-SPM variants with DM-OFDM-IM, GDM-OFDM-IM and MM-OFDM-IM.}
		\label{fig:fig2}
\end{figure}

In Fig.~\ref{fig:fig2}, we compare the data rates of the proposed OFDM-SPM schemes with DM-OFDM-IM, GDM-OFDM-IM, and MM-OFDM-IM. The data rate results in terms of the number of index bits per subcarrier are given as a function of $N$. Since all the schemes considered in this figure activate all subcarriers and the modulation order of the carried symbols on each subcarrier can be chosen to be the same, we ignore the modulation bits transmitted per subcarrier. Also, we do not restrict the codebook sizes to a power of two, since all the codewords in a codebook can be utilized by a binary coding technique such as Huffman coding regardless of the number of codewords \cite{Wang2017,Coon2019}. To reach the maximum number of index bits, DM-OFDM-IM  modulates  half of the subcarriers by one of the distinguishable constellations and the other half by the other constellation. Moreover, we assume $\mathcal{K}=\big\{0, 1, \ldots, N\big\}$ for GDM-OFDM-IM. To reach the maximum numbers of set partitions for OFDM-SPM and OFDM-OSPM, we substitute $ K_N \in \big\{\floor{e^{W(N)}-1}, \ceil{e^{W(N)}-1} \big\}$ and $\breve{K}_N=\frac{N}{2 \ln2}$, respectively, into $K$.  The data rate results verify the remarks in the previous section and indicate that all SPM variants outperform DM-OFDM-IM and GDM-OFDM-IM  for most of the values of $N$. On the other hand, although MM-OFDM-IM considerably outperforms OFDM-SPM and OFDM-FSPM, it is outperformed by OFDM-OSPM and OFDM-OFSPM for almost all values of $N$.

\begin{figure}[t]
		\centering
        \includegraphics[width=8.5cm]{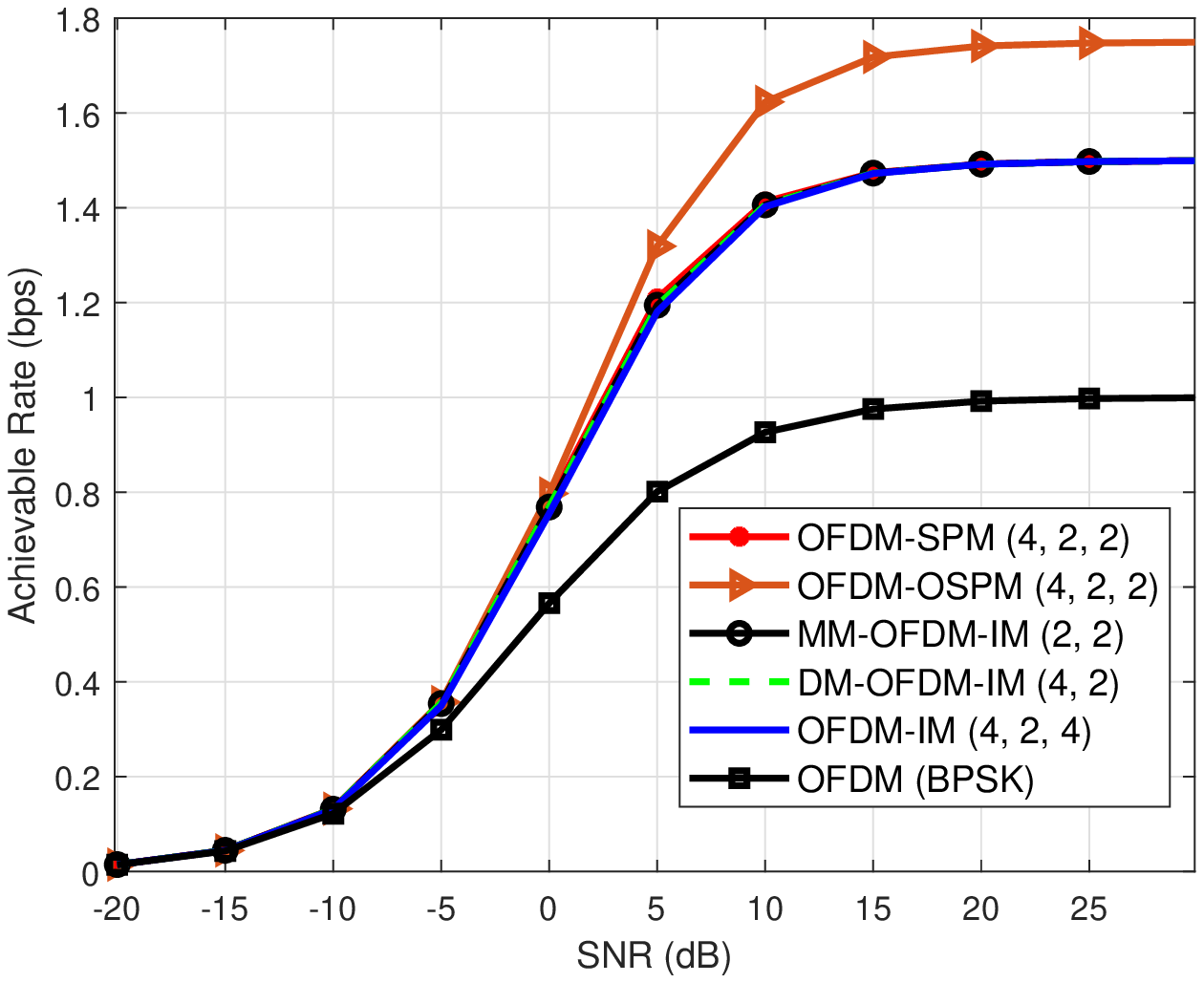}
		\caption{Achievable rate comparison of OFDM-SPM $(4,2,2)$ and OFDM-OSPM $(4,2,2)$  with MM-OFDM-IM $(2,2)$, DM-OFDM-IM $(4,2)$, OFDM-IM $(4,2,4)$ and OFDM (BPSK).}
		\label{fig:fig2a}
\end{figure}

To gain more insight into the  achievable rate of the proposed schemes, one can substitute a finite input symbol set for such schemes into the standard mutual information expression. Assuming equally likely codewords and applying the same approach as discussed in \cite{Ishikawa2016,Ng2006}, the achievable rate of the proposed schemes can be computed  by evaluating
\begin{align}\label{eq:eq10}
    R=\frac{1}{N}\bigg(f-\frac{1}{2^f}\sum_{i=1}^{2^f}\operatorname{E}_{\textbf{h},\textbf{n}}\bigg[\log_2\sum_{j=1}^{2^f}e^{\delta(i,j)}\bigg]\bigg)
\end{align}
where $\delta(i,j)=\frac{-||\text{diag}(\textbf{h})(\textbf{x}^i-\textbf{x}^j)+\textbf{n}||^2+||\textbf{n}||^2}{N_0}$. Although this approach does not yield a closed-form solution, we can easily obtain numerical data for the theoretical achievable rate of the proposed schemes.

\begin{figure}[t]
		\centering
	    \includegraphics[width=8.5cm]{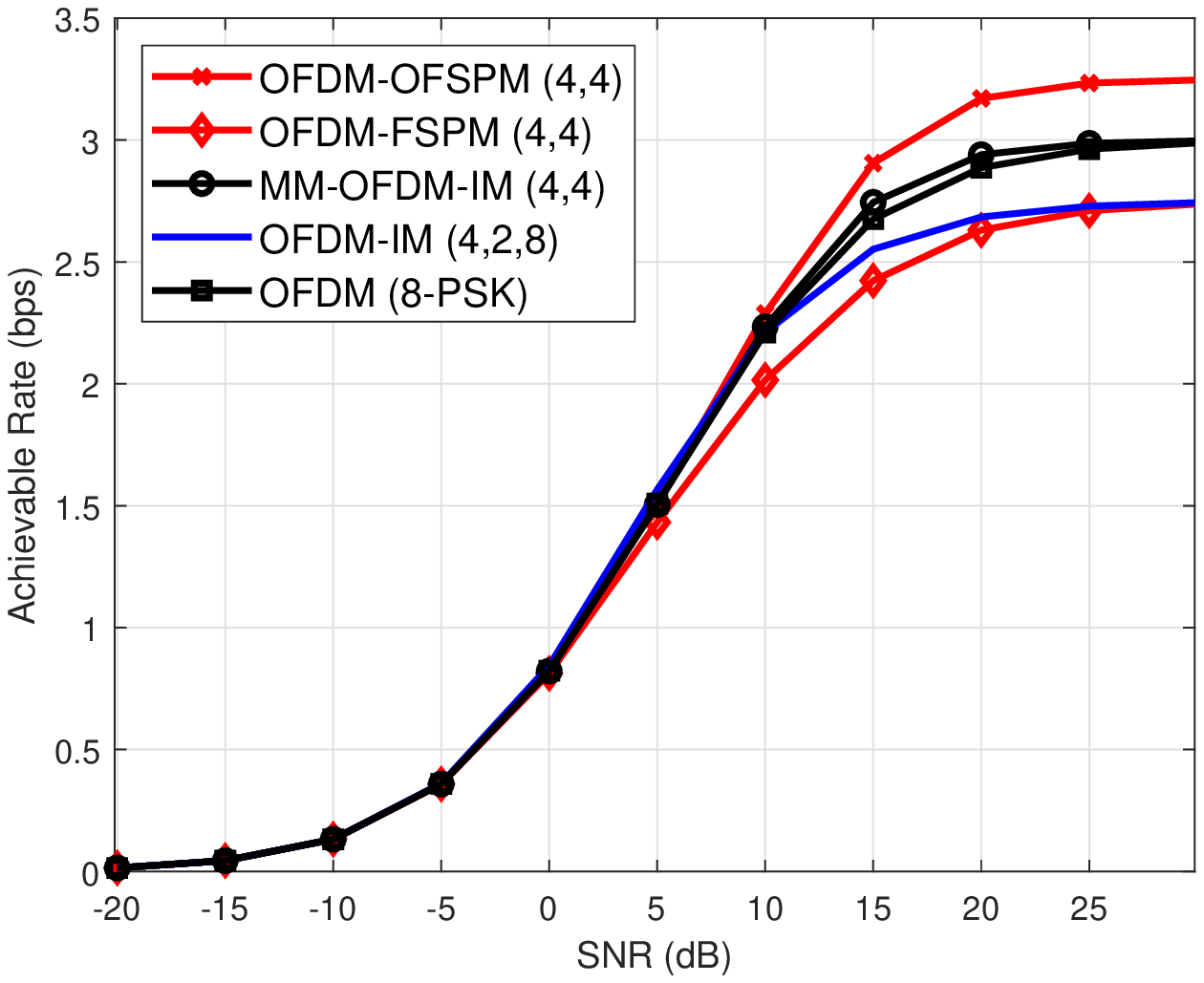}
		\caption{Achievable rate comparison of OFDM-OFSPM $(4,4)$ and OFDM-FSPM $(4,4)$  with MM-OFDM-IM $(4,4)$, OFDM-IM $(4,2,8)$ and OFDM (8-PSK).}
		\label{fig:fig22}
	\end{figure}

We compare the achievable rates of OFDM-SPM variants with the achievable rates of MM-OFDM-IM, DM-OFDM, OFDM-IM and OFDM schemes in Figs. \ref{fig:fig2a} and \ref{fig:fig22}. The achievable rate curves provided in these figures are obtained by using \eqref{eq:eq10}. The results in Fig. \ref{fig:fig2a} are given for OFDM-SPM $(4,2,2)$, OFDM-OSPM $(4, 2, 2)$, MM-OFDM-IM $(2, 2)$, DM-OFDM-IM $(4, 2)$, OFDM-IM $(4, 2, 4)$ and OFDM (BPSK) schemes. Here, OFDM-SPM $(4, 2, 2)$  and OFDM-OSPM $(4, 2, 2)$ encoders produce $\stirlingii{4}{2}=7$ and $2!\stirlingii{4}{2}=14$ codewords, and then $2^{\log_2 \floor{7}}=4$ and $2^{\log_2 \floor{14}}=8$ set partitions are selected by Algorithm \ref{alg:alg1}. As can be observed from the figure, OFDM-SPM $(4,2,2)$, MM-OFDM-IM $(2,2)$, DM-OFDM-IM $(4,2)$ and OFDM-IM $(4,2,4)$ exhibit almost the same achievable rate performance for all SNR values. On the other hand, OFDM-OSPM $(4,2,2)$ outperforms all other IM schemes. Moreover, to achieve 1.5 bps, the SNR requirement for OFDM-OSPM $(4,2,2)$ is approximately 12 dB lower than for other IM schemes.

In Fig. \ref{fig:fig22}, we compare the achievable rates of OFDM-OFSPM $(4, 4)$ and OFDM-FSPM $(4, 4)$  with MM-OFDM-IM $(4, 4)$, OFDM-IM $(4, 2, 8)$ and OFDM (8-PSK). Here, OFDM-FSPM $(4, 4)$  and OFDM-OFSPM $(4, 4)$ encoders produce $B_N=15$ and $\Breve{B}_N=75$ codewords, respectively. 32 codewords are then selected according to Algorithm \ref{alg:alg2} for  OFDM-OFSPM $(4, 4)$ in a way that the rank of the difference matrices for the codeword pairs in the selected codebook is at least two \cite{Wen2017}. For OFDM-FSPM $(4, 4)$, Algorithms \ref{alg:alg1} and \ref{alg:alg2} cannot return $2^{\floor{\log_2 15}}=8$ codewords; they return only four and five codewords, respectively. To fully exploit the codewords generated by this scheme and to achieve a high data rate at high SNR, we use three other codewords in the OFDM-FSPM (4, 4) scheme in addition to the codewords generated by Algorithm \ref{alg:alg2}. This limits the minimum Hamming distance of the codeword pairs to one. This explains why the OFDM-FSPM (4, 4) scheme is outperformed by all other schemes. However, Fig. \ref{fig:fig22} shows that  OFDM-OFSPM $(4,4)$  is capable of providing a better achievable rate than all other schemes at high SNR. Moreover, OFDM-OFSPM $(4,4)$  is, theoretically, able to achieve 2 bps with an SNR requirement that is almost 13 dB lower than that of MM-OFDM-IM $(4,4)$.

\subsection{Bit-error Rate}
In this subsection, we compare the proposed OFDM-SPM schemes with conventional OFDM and IM benchmarks in terms of BER performance.      

\begin{figure}[t]
		\centering
		\includegraphics[width=8.5cm]{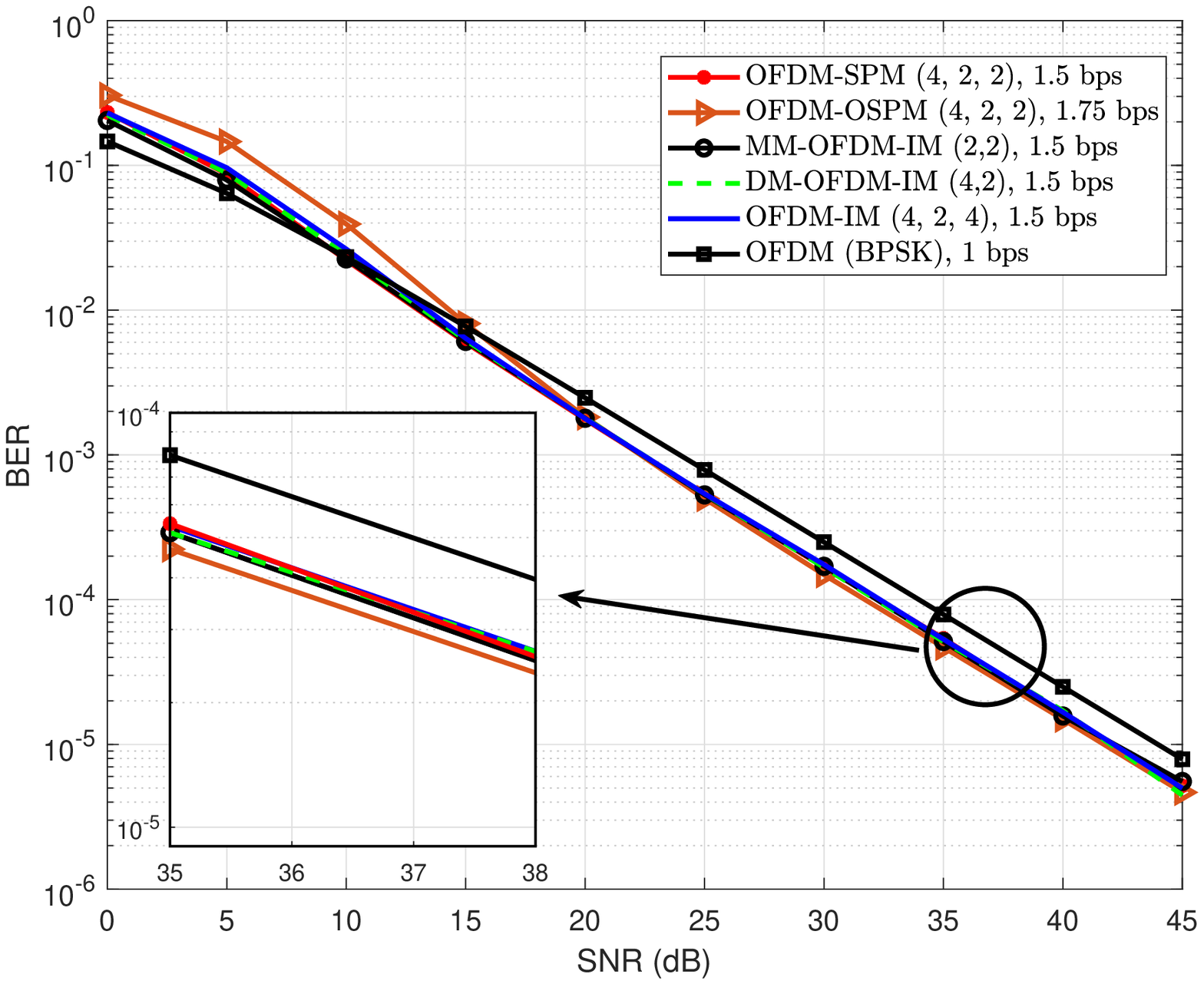}
		\caption{BER comparison of OFDM-SPM $(4,2,2)$ and OFDM-OSPM $(4,2,2)$ with MM-OFDM-IM $(2,2)$, DM-OFDM-IM $(4,2)$, OFDM-IM $(4,2,4)$ and OFDM (BPSK).}
		\label{fig:fig3}
\end{figure}

In Fig.~\ref{fig:fig3}, we compare the BER performance of OFDM-SPM $(4, 2, 2)$  and OFDM-OSPM $(4, 2, 2)$ with MM-OFDM-IM $(2, 2)$, DM-OFDM-IM $(4,2)$, OFDM-IM $(4, 2, 4)$,  and conventional OFDM (BPSK). Except for OFDM-OSPM $(4, 2, 2)$ and conventional OFDM (BPSK), all schemes exhibit the same spectral efficiency, which is 1.5 bps. Spectral efficiencies for OFDM-OSPM $(4, 2, 2)$ and conventional OFDM (BPSK) are 1.75 bps and 1 bps, respectively. Here, OFDM-SPM $(4, 2, 2)$  and OFDM-OSPM $(4, 2, 2)$ encoders produce $\stirlingii{4}{2}=7$ and $2!\stirlingii{4}{2}=14$ codewords, and then $2^{\log_2 \floor{7}}=4$ and $2^{\log_2 \floor{14}}=8$ set partitions are selected by Algorithm \ref{alg:alg1}. As seen from the figure, OFDM-SPM $(4, 2, 2)$ exhibits almost the same BER performance as OFDM-IM $(4, 2, 4)$, DM-OFDM-IM $(4,2)$, and MM-OFDM-IM $(2, 2)$, and it outperforms conventional OFDM (BPSK) at medium-to-high SNR. More importantly, although OFDM-OSPM $(4, 2, 2)$ has higher spectral efficiency than other schemes, it provides marginally better BER performance compared to other schemes at high SNR.  The reason behind this is the codewords related to set partitions in the OFDM-OFSPM scheme have the capability of preserving the same Hamming distance properties as the codewords related to the permutations in the MM-OFDM-IM scheme or the combinations in the DM-OFDM-IM scheme, and the number of such codewords in the OFDM-OFSPM scheme is larger than those in the DM-OFDM-IM and MM-OFDM-IM schemes. Having a higher number of these codeweords is desirable in the high SNR since they are capable of introducing diversity order of two unlike the codewords related to conventional modulation, which limit the diversity order to one. Therefore, they mitigate the effect of the codewords related to conventional modulation on the BER performance.

\begin{figure}[t]
		\centering
		\includegraphics[width=8.5cm]{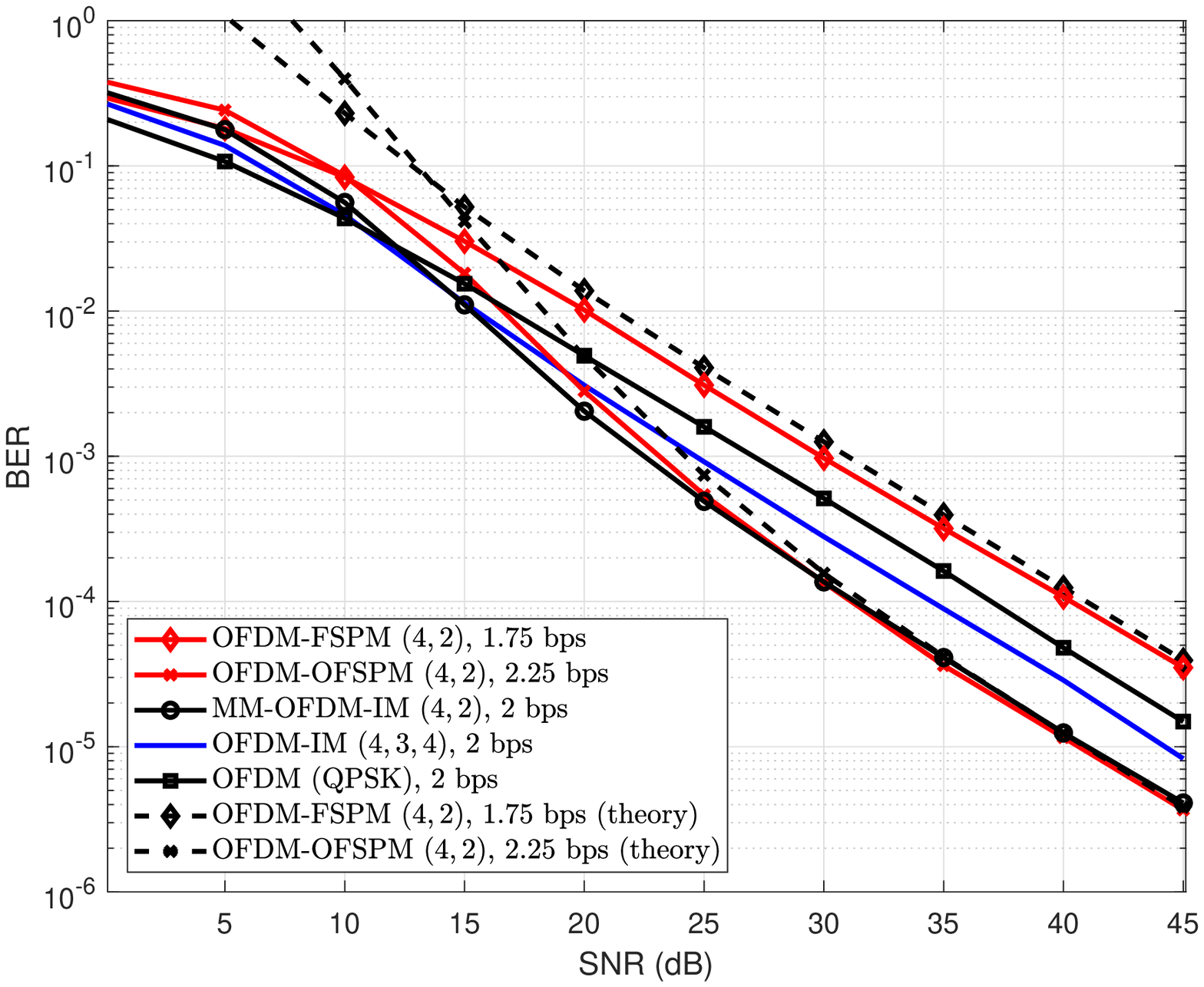}
		\caption{BER comparison of OFDM-FSPM $(4, 2)$ and OFDM-OFSPM $(4, 2)$ with MM-OFDM-IM $(4,2)$, OFDM-IM $(4,3,4)$ and OFDM (QPSK).}
		\label{fig:fig4}
\end{figure}

In Fig.~\ref{fig:fig4},   OFDM-FSPM $(4, 2)$  and OFDM-OFSPM $(4, 2)$  are compared with MM-OFDM-IM $(4, 2)$, OFDM-IM $(4, 3, 4)$, and conventional OFDM (QPSK). Except for OFDM-FSPM $(4, 2)$  and OFDM-OFSPM $(4, 2)$, all schemes have the same spectral efficiency of 2 bps. Spectral efficiencies for OFDM-FSPM $(4, 2)$  and OFDM-OFSPM $(4, 2)$ are 1.75 bps and 2.25 bps, respectively. Here, OFDM-FSPM $(4, 2)$  and OFDM-OFSPM $(4, 2)$ encoders produce $B_N=15$ and $\Breve{B}_N=75$ codewords. Then, 32 codewords are selected according to Algorithm \ref{alg:alg2} for OFDM-OFSPM $(4, 2)$.  As discussed in the previous section, Algorithms \ref{alg:alg1} and \ref{alg:alg2} are not capable of providing $2^{\floor{\log_2 15}}=8$ codewords for OFDM-FSPM $(4, 2)$. However, we use eight codewords of OFDM-FSPM $(4, 2)$, including the five FSPM codewords provided by Algorithm \ref{alg:alg2}, to achieve a higher data rate.  We also provide results on the theoretical upper-bound for the OFDM-SPM schemes. As observed from the figure, upper-bound curves are consistent with computer simulations, especially at high SNR. Although OFDM-FSPM $(4, 2)$ cannot provide a BER advantage relative to OFDM-IM $(4, 3, 4)$ and MM-OFDM-IM $(4, 2)$, OFDM-OFSPM $(4, 2)$ exhibits superior BER performance relative to all benchmarks at high SNR while achieving enhanced spectral efficiency. These results arise from the fact that the set partitions in the selected OFDM-FSPM codebook exhibit lower rank than the set partitions in the selected OFDM-OFSPM codebook. Moreover, the codewords related to set partitions in OFDM-OFSPM  are capable of preserving the same minimum rank property as the codewords related to the permutations in the MM-OFDM-IM scheme, although the number of index bits for the former increases.

\begin{figure}[t]
		\centering
		\includegraphics[width=8.5cm]{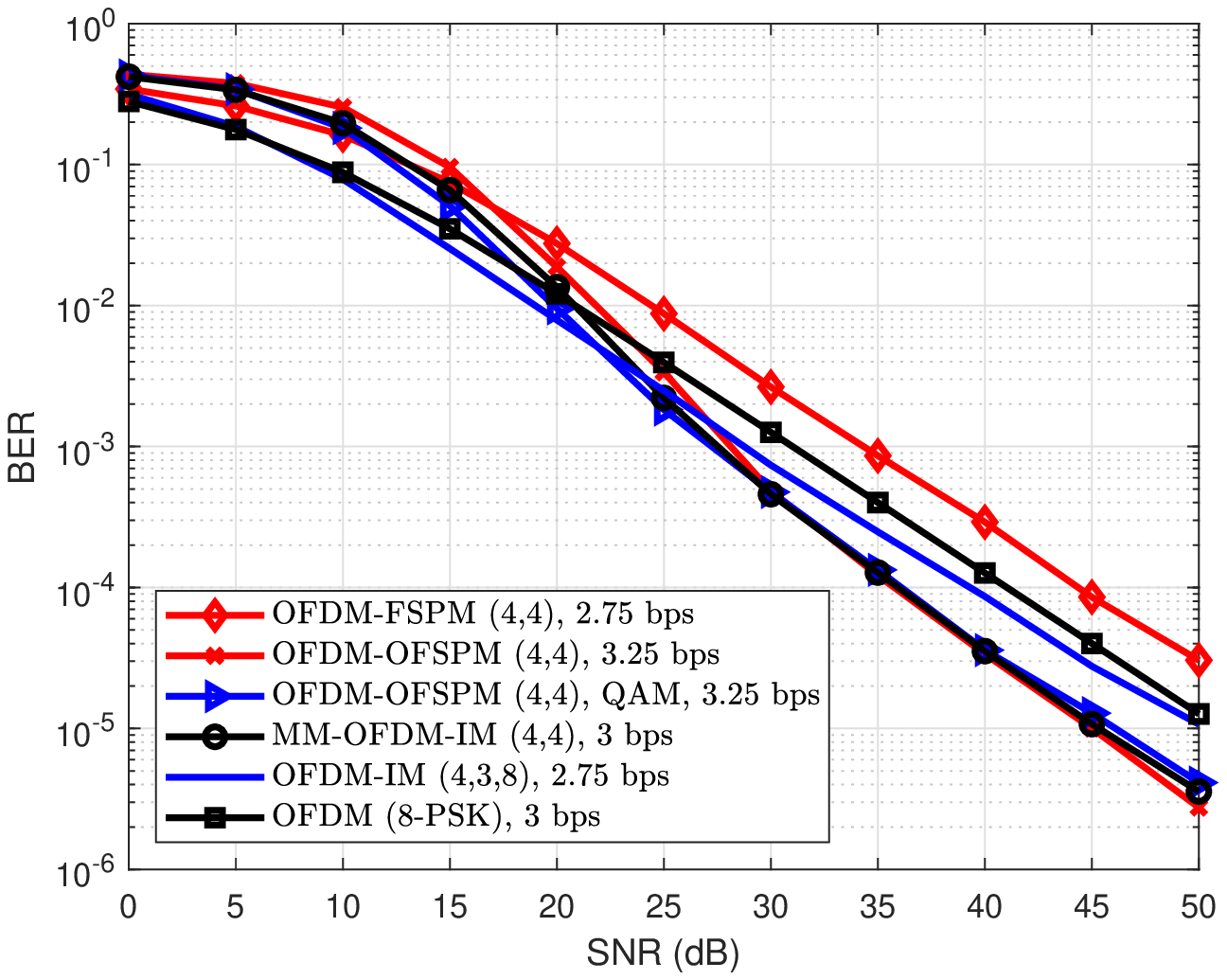}
		\caption{BER comparison of OFDM-FSPM $(4, 4)$ and OFDM-OFSPM $(4, 4)$ with MM-OFDM-IM $(4,4)$, OFDM-IM $(4,3,8)$ and OFDM (8-PSK).}
		\label{fig:fig4last}
\end{figure}

In Fig. \ref{fig:fig4last}, we compare the BER performance of  OFDM-FSPM $(4, 4)$  and OFDM-OFSPM $(4, 4)$ with MM-OFDM-IM $(4, 4)$, OFDM-IM $(4, 3, 8)$, and conventional OFDM (8-PSK)  schemes. We also provide a BER curve (``OFDM-OFSPM $(4, 4)$, QAM'') for OFDM-OFSPM employing four different 4-QAM constellations to distinguish the OFDM-OFSPM codewords. These constellations are obtained by employing the set partitioning technique described in \cite{Ungerboeck1982} to a 16-QAM constellation. We use the same SPM codewords as Fig. \ref{fig:fig4} for OFDM-SPM variants.  In this figure, OFDM-FSPM $(4, 4)$ and OFDM-IM $(4, 3, 8)$; MM-OFDM-IM $(4, 4)$ and OFDM (8-PSK); OFDM-OFSPM $(4, 4)$ and OFDM-OFSPM $(4, 4)$, QAM schemes have the same spectral efficiencies which are 2.75 bps, 3 bps, and 3.25 bps, respectively. Similar results to Fig. \ref{fig:fig4} can be observed in Fig. \ref{fig:fig4last} for a higher modulation order. On the other hand, OFDM-OFSPM $(4, 4)$, QAM outperforms both OFDM-OFSPM $(4, 4)$ and MM-OFDM-IM $(4, 4)$ at low SNR values; however, OFDM-OFSPM $(4, 4)$ exhibits slightly better BER performance than OFDM-OFSPM $(4, 4)$, QAM and MM-OFDM-IM $(4, 4)$ at high SNR values. The behavior of the BER curves at low SNR values can be explained by the minimum Euclidean distances between the codewords. The minimum Euclidean distance between ``OFDM-OFSPM $(4, 4)$, QAM'' codewords is 0.8944; however, the minimum Euclidean distance between OFDM-OFSPM $(4, 4)$ and  MM-OFDM-IM $(4, 4)$  codewords is just 0.5518. On the other hand, the BER performance at high SNR will be dominated by the Euclidean distances between the codewords whose difference matrix has the minimum rank. In that case, the Euclidean distance between OFDM-OFSPM $(4, 4)$ and  MM-OFDM-IM $(4, 4)$  codewords is given by 1.4142; however, the Euclidean distance between ``OFDM-OFSPM $(4, 4)$, QAM'' codewords is just 1.2649. The comparison between OFDM-OFSPM $(4, 4)$ and  MM-OFDM-IM $(4, 4)$ results from the number of index bits. OFDM-OFSPM $(4, 4)$ produces more index bits than MM-OFDM-IM $(4, 4)$ by using set partitions rather than permutations. Such index bits become undesirable at low SNR due to the minimum Euclidean distance between the associated codewords. However, at high SNR, they become desirable since these codewords provide better Hamming distance properties than the codewords associated with conventional modulation bits.               

\section{Conclusion}\label{section:section6}
In this paper, we proposed a novel modulation concept, which we call set partition modulation. We represented several variants of the concept and showed a practical implementation in the context of OFDM. Moreover, we defined a codebook selection problem for the proposed techniques and expressed such a problem as a clique problem in graph theory. We further provided an efficient solution for the codebook selection. Then, we investigated the performance of the new techniques in terms of their data rates and BER characteristics, and we presented asymptotic results regarding the performance. We compared the proposed OFDM-SPM variants with the appropriate benchmarks. Through computer simulations and theoretical calculations, it is shown that the proposed SPM schemes can provide noteworthy improvements compared to conventional OFDM, OFDM-IM, DM-OFDM-IM, and MM-OFDM-IM in terms of data rate and BER.  


 \bibliography{Journal_draft}

\end{document}